# Advancements in UAV-based Integrated Sensing and Communication: A Comprehensive Survey

Manzoor Ahmed, Ali Arshad Nasir, Mudassir Masood, Kamran Ali Memon, Khurram Karim Qureshi, Feroz Khan, Wali Ullah Khan, Fang Xu, and Zhu Han

*Abstract*—Unmanned aerial vehicle (UAV)-based integrated sensing and communication (ISAC) systems are poised to revolutionize next-generation wireless networks by enabling simultaneous sensing and communication (S&C). This survey comprehensively reviews UAV-ISAC systems, highlighting foundational concepts, key advancements, and future research directions. We explore recent advancements in UAV-based ISAC systems from various perspectives and objectives, including advanced channel estimation (CE), beam tracking, and system throughput optimization under joint sensing and communication S&C constraints. Additionally, we examine weighted sum rate (WSR) and sensing trade-offs, delay and age of information (AoI) minimization, energy efficiency (EE), and security enhancement. These applications highlight the potential of UAV-based ISAC systems to improve spectrum utilization, enhance communication reliability, reduce latency, and optimize energy consumption across diverse domains, including smart cities, disaster relief, and defense operations. The survey also features summary tables for comparative analysis of existing methodologies, emphasizing performance, limitations, and effectiveness in addressing various challenges. By synthesizing recent advancements and identifying open research challenges, this survey aims to be a valuable resource for developing efficient, adaptive, and secure UAV-based ISAC systems.

## I. Introduction

Integrated sensing and communication (ISAC) has emerged as a transformative technology central to next-generation wireless systems, particularly sixth-generation (6G) networks [1], [2]. By merging radar sensing with wireless communication, ISAC overcomes the inefficiencies of traditional standalone systems. It leverages shared resources such as hardware, spectrum, and energy to boost operational efficiency, reduce costs, and enhance sustainability [3], [4]. In contrast, traditional systems necessitate separate devices for sensing and communication (S&C), resulting in higher energy use, increased spectral demands, and significant deployment expenses [5]. ISAC resolves these issues by integrating S&C functionalities, leading to improved energy efficiency (EE), enhanced spectral efficiency (SE), and minimized hardware redundancy [6], [7]. At the core of ISAC lies the synergy between S&C. This interaction enhances functionalities: sensing for communication (S4C) significantly improves communication quality by providing real-time environmental insights, such as obstacle locations and channel state information (CSI) essential for reliable connectivity. Conversely, communication for sensing (C4S) leverages existing communication infrastructure for distributed sensing tasks, including data sharing, fusion, and coordination [8]. This interplay optimizes capabilities and unlocks new design opportunities, fostering innovative applications and more efficient resource use [9], [10].

UAV-based ISAC systems represent a promising implementation, leveraging UAVs' unique capabilities to enhance integrated S&C [8]. They enable simultaneous S&C by sharing resources like spectrum and hardware, overcoming traditional ground-based limitations [11]. Unlike fixed infrastructure, UAVs can dynamically navigate three-dimensional (3D) space, enabling them to circumvent obstacles that hinder line-of-sight (LoS) links [12]. This adaptability guarantees reliable communication and precise sensing in complex environments—urban areas, disaster zones, and remote locations—where fixed infrastructure is lacking or inadequate, as illustrated in Fig. 1.

UAV mobility significantly enhances the adaptability of ISAC systems. By facilitating real-time trajectory optimization, UAVs enable dynamic position adjustments in response to changing environmental conditions. This capability not only optimizes S&C tasks but also maximizes overall performance [5], [13]. Such adaptability introduces new degrees of freedom (DoFs) in system design, positioning UAV-ISAC systems as ideal solutions for high-responsiveness applications. These applications include intelligent communication networks, autonomous systems, emergency relief efforts, surveillance, and public safety initiatives [8], [14]. In the context of disaster management, UAVs can navigate through debris to establish robust communication links.

UAV-ISAC systems combine adaptability and flexibility to tackle essential performance trade-offs in modern wireless networks [14]. These trade-offs encompass several key factors: balancing the weighted sum rate (WSR) with sensing capabilities, optimizing throughput while adhering to joint

Manzoor Ahmed and Fang Xu are with the School of Computer and Information Science and also with Institute for AI Industrial Technology Research, Hubei Engineering University, Xiaogan City, 432000, China (e-mails: manzoor.achakzai@gmail.com, xf@hbeu.edu.cn).

Ali Arshad Nasir, Mudassir Masood, Khurram Karim Qureshi are with the Interdisciplinary Research Center for Communication Systems and Sensing, King Fahd University of Petroleum and Minerals, Dhahran, 31261, Saudi Arabia and Department of Electrical Engineering, King Fahd University of Petroleum and Minerals, Dhahran, 31261, Saudi Arabia. (e-mails:anasir@kfupm.edu.sa; mudassir@kfupm.edu.sa; kqureshi@kfupm.edu.sa ).

Kamran Ali Memon is with the Interdisciplinary Research Center for Communication Systems and Sensing, King Fahd University of Petroleum and Minerals, Dhahran, 31261, Saudi Arabia (e-mail: ali.kamran@kfupm.edu.sa).

Feroz Khan is with the School of Electronic Engineering, Beijing University of Posts and Telecommunications, Beijing, China (e-mail: ferozkhan687@gmail.com).

Wali Ullah Khan is with the Interdisciplinary Centre for Security, Reliability, and Trust (SnT), University of Luxembourg, 1855 Luxembourg City, Luxembourg (e-mails: waliullah.khan@uni.lu).

Zhu Han is with the Department of Electrical and Computer Engineering at the University of Houston, Houston, TX 77004 USA, and also with the Department of Computer Science and Engineering, Kyung Hee University, Seoul, South Korea, 446-701. (e-mail: hanzhu22@gmail.com).

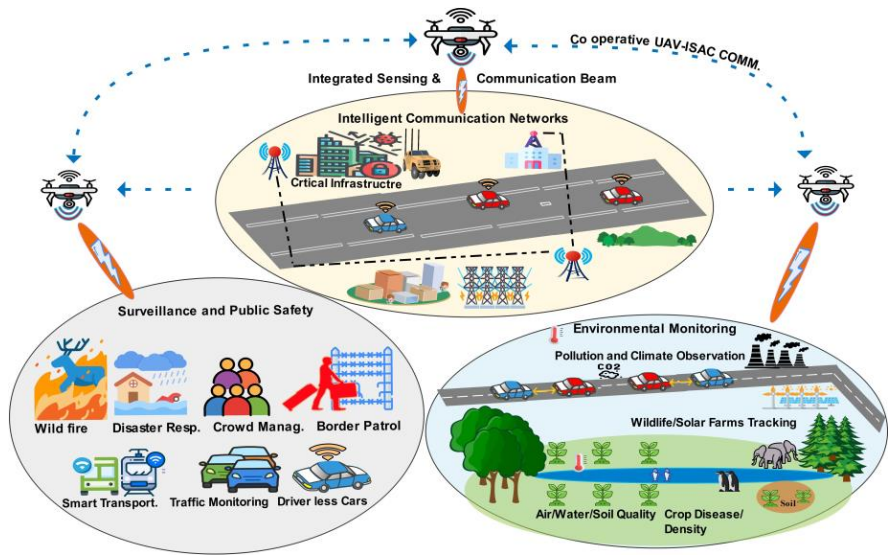

Fig. 1: Applications of UAV-based ISAC systems in diverse scenarios

TABLE I: Summary of Important Acronyms

| Abbreviation | Definition |
| --- | --- |
| WSR | Weighted Sum Rate |
| UAVs | Unmanned Aerial Vehicles |
| THz | Terahertz |
| S&C | Sensing and Communication |
| SR | Sum Rate |
| SOP | Secrecy Outage Probability |
| SNR | Signal to Noise Ratio |
| SDR | Semidefinite Relaxation |
| SCE | Secure Computation Efficiency |
| SCA | Successive Convex Approximation |
| RIS | Reconfigurable Intelligent Surfaces |
| PPO | Proximal Policy Optimization |
| PSO | Particle Swarm Optimization |
| PLS | Physical Layer Security |
| NTNs | Non-Terrestrial Networks |
| NLoS | Non-Line-of-Sight |
| NOMA | Non-Orthogonal Multiple Access |
| ML | Machine Learning |
| MEC | Mobile Edge Computing |
| LoS | Line of Sight |
| ISAC | Integrated Sensing and Communication |
| IoT | Internet of Things |
| GUE | Ground User Equipment |
| CSI | Channel State Information |
| CRB | Cramér–Rao Bounds |
| CE | Channel Estimation |
| BS | Base Station |
| AI | Artificial Intelligence |
| 6G | Sixth Generation |
| Eve | Eavesdropper |
| DDPG | Deep Deterministic Policy Gradient |
| DRL | Deep Reinforcement Learning |
| AO | Alternate Optimization |
| AoI | Age of Information |

sensing and communications (S&C) constraints, minimizing latency and AoI, enhancing EE, and ensuring secure operations within dynamic environments. Dynamic position adjustments effectively mitigate threats such as interference and eavesdropping. This capability is essential for ensuring robust and secure communication in reliability-critical applications, including military surveillance, disaster recovery, and various other operations [15].

This survey reviews advancements in UAV-based ISAC systems, focusing on fundamental principles, operational capabilities, and diverse applications. It explores recent advancements in UAV-based ISAC systems from various perspectives and objectives, such as advanced channel estimation (CE), target and beam tracking, throughput optimization, WSR, and sensing. Furthermore, it addresses critical factors including delay and AoI, EE, and security. In addition, summary tables outline various optimization strategies for each UAV-ISAC advancement, facilitating a comparative analysis of methodologies. The analysis highlights the strengths and limitations of different approaches while offering insights into their applicability in varying scenarios. By synthesizing recent developments and identifying key research challenges, this survey serves as a roadmap for creating efficient, adaptive, and secure UAV-based ISAC systems. With applications that span smart cities, disaster recovery, defense, and autonomous operations, these systems have the potential to revolutionize wireless networks and provide innovative solutions for a connected, data-driven world.

*A. Motivation*

The increasing demands of 6G wireless networks underscore the necessity for advanced solutions that unify S&C within a cohesive framework. In this regard, ISAC has emerged as a transformative approach, mitigating the inefficiencies of standalone systems by leveraging shared resources such as hardware, spectrum, and energy. However, conventional ISAC systems encounter challenges in complex environments where fixed infrastructure is obstructed or absent. This issue is particularly prevalent in urban settings, disaster areas, and remote regions. In response, UAV-based ISAC systems present a promising alternative, adeptly addressing the limitations of traditional infrastructures. Consequently, validating UAV-based ISAC systems as an effective solution highlights the capability of UAVs to navigate and overcome the challenges posed by conventional infrastructure. In this context, UAV-based ISAC systems stand out as a compelling solution, skillfully navigating the constraints imposed by traditional infrastructures.

Integrating the flexibility and mobility of UAVs with the synergistic capabilities of ISAC significantly enhances reliable line-of-sight (LoS) communication, enables real-time sensing, and supports dynamic trajectory optimization. UAV-ISAC systems demonstrate exceptional adaptability, which makes them highly effective across diverse applications. However, despite their significant potential, several unresolved challenges remain. This survey aims to address these gaps by providing a comprehensive review of advancements in UAV-ISAC technology and suggesting future research directions.

*B. Related Surveys*

Table II highlights ISAC as a transformative technology with significant potential for next-generation wireless networks. Surveys have examined ISAC's advancements, challenges, and applications, offering key insights into its capabilities. For example, ISAC's crucial role in the Internet of Things (IoT) underscores its importance in achieving seamless integration and improved system performance [16]. Several studies have investigated the theoretical limits of ISAC systems, delving into the intricate trade-offs between sensing and communication. Additionally, these studies have identified optimization strategies aimed at effectively balancing these two domains [17]. Moreover, ISAC waveform design strategies have been systematically classified into communication-centric, sensing-centric, and joint optimization approaches. These classifications propose innovative pathways for the evolution of next-generation ISAC systems [18].

ISAC's application in vehicular networks has emerged as a pivotal area of research, effectively addressing essential challenges such as low-latency communication, accurate localization, and advancements in autonomous driving. The insights gained from these studies reinforce the necessity for integrated solutions within future transportation systems, ultimately leading to improvements in road safety, enhanced communication efficiency, and increased sensing accuracy [19], [20]. Furthermore, the rise of UAV-enabled ISAC systems has garnered considerable attention, primarily due to their mobility, flexibility, and real-time adaptability. This positions them as exceptional companions for navigating complex and dynamic environments [21].

The research on ISAC in 5G and 6G networks has emerged as a crucial area of focus, particularly regarding signal design and optimization. Recent studies have focused on waveform optimization and the development of energy-efficient protocols, which are essential for significantly enhancing overall network performance [22]. Furthermore, machine learning (ML) techniques are increasingly being harnessed to drive data-driven improvements in critical areas such as localization, beamforming, and sensing optimization, thereby addressing the evolving demands of modern communication networks. In addition, ML plays a vital role in enhancing resource allocation and improving system efficiency within ISAC networks [23], [24]. Artificial intelligence (AI) also emerges as a central theme in ISAC research, fostering innovations in system scalability, resource optimization, and real-time adaptability [25].

Research on ISAC in 6G networks has highlighted several critical challenges, including interference management, spectrum sharing, and the inherent nonstationarity of the environment. To address these challenges and enhance operational efficiency, advanced techniques such as beamforming and hybrid systems are being proposed [26]. Further, the integration of ISAC into smart cities, autonomous vehicles, and other emerging applications demonstrates its potential in next-generation use cases [27]. Another area of growing interest is the role of intelligent metasurfaces in enabling ISAC systems. These metasurfaces can dynamically manipulate electromagnetic waves, improving SE, adaptability, and overall system performance [28]. Furthermore, the seamless integration of sensing, communication, and computation has been emphasized as a crucial factor for real-time resource management and coordination in 6G applications [29].

To effectively tackle interference in dense and dynamic environments, researchers have proposed innovative interference management strategies that significantly enhance the performance of ISAC systems [30]. In this context, orthogonal time-frequency space (OTFS) modulation has emerged as a promising technique, providing improved SE and performance in complex scenarios [31]. Moreover, security and privacy remain significant concerns in ISAC systems. As a response, researchers are focusing on developing strategies to counter potential threats and improve the robustness of these integrated systems [32]. Additionally, the potential of terahertz (THz) frequencies for ISAC has been explored. This research emphasizes overcoming challenges such as high path loss and beam management to fully leverage their capabilities [33]. Together, these studies offer a comprehensive understanding of ISAC's potential, challenges, and future opportunities, laying a strong foundation for its implementation in advanced wireless communication networks.

*C. Contributions*

In contrast to the existing surveys in Table II, our survey offers a comprehensive examination of UAV-based ISAC systems. It highlights the following key contributions:

- **Comprehensive Foundation:** This section provides a comprehensive overview of UAVs and ISAC, emphasizing the significance of their integration in wireless networks.
- **Detailed Examination of Advancements:** Investigates real-world applications of UAV-based ISAC technologies, focusing on critical functionalities such as CE, beam/target tracking, maximizing system throughput, balancing WSR with sensing requirements, minimizing delays and AoI, improving EE, and enhancing security. Summary tables outline UAV roles, sensing metrics, methodologies, and comparisons of strategies.
- **Lessons Learned:** This survey underscores essential insights from UAV-based ISAC systems, emphasizing the significance of optimal UAV placement and trajectory design to maximize throughput, coverage, and sensing

TABLE II: Survey Papers' Comparison Related to UAV-based ISAC

| Ref. | Year | Focus of the Related Surveys | CE/Target and Beam Tracking | System Throughput | WSR and Sensing | Delay | EE | Security |
|---|---|---|---|---|---|---|---|---|
| [16] | 2021 | Highlights ISAC's role in enabling seamless IoT applications | × | × | × | × | × | × |
| [17] | 2022 | Explores the theoretical trade-offs and performance limits of ISAC systems | × | × | × | × | × | × |
| [18] | 2022 | Explores joint waveform design to balance ISAC sensing and communication requirements | × | × | × | × | × | × |
| [19] | 2022 | Discusses ISAC's potential in enabling advanced vehicular networks | × | × | × | × | × | × |
| [20] | 2023 | Highlights ISAC as a solution in CAV networks' challenges, focusing on its application and advancements for 6G | × | × | × | × | × | × |
| [21] | 2023 | Highlights UAVs' role in ISAC for real-time adaptability and sensing flexibility | * | × | × | × | * | × |
| [22] | 2023 | Focuses on signal design and optimization for ISAC in 5G-A and 6G | × | × | × | × | × | × |
| [23] | 2023 | Highlights data-driven approaches for enhancing ISAC efficiency and accuracy | * | × | × | × | * | × |
| [24] | 2024 | Highlights machine learning's role in improving ISAC sensing, communication, and resource allocation. | * | × | × | × | * | × |
| [25] | 2024 | Explores the advancements from the collaboration of sensing, communication, and AI | * | × | × | × | × | × |
| [26] | 2024 | Focuses on channel modeling approaches and opportunities for ISAC in 6G networks | *** | × | × | × | × | × |
| [27] | 2024 | Outlines a vision for ISAC networks and explores how signal processing, optimization, and ML techniques can be harnessed to realize their potential within the 6G framework. | *** | × | × | × | * | × |
| [28] | 2024 | Highlights the role of metasurfaces in enhancing ISAC efficiency and adaptability | * | × | * | × | * | * |
| [29] | 2024 | Examines the integration of sensing, communication, and computation in next-generation networks | * | * | × | × | * | ** |
| [30] | 2024 | Discusses strategies for mitigating interference in ISAC systems | × | × | × | × | × | × |
| [31] | 2024 | Highlights OTFS modulation's role in improving ISAC SE | × | × | × | × | × | × |
| [32] | 2024 | Focuses on enhancing security and privacy in ISAC systems | × | × | × | × | × | *** |
| [33] | 2024 | Highlights the potential and challenges of using THz frequencies for ISAC | × | × | × | × | × | × |
| Our | 2025 | This comprehensive survey on UAV-based ISAC systems focuses on their ability to simultaneously perform S&C. It highlights key topics such as CE, beam tracking, system throughput optimization, WSR and sensing, EE, and security. Additionally, it compares existing methods, emphasizes lessons learned, and identifies future research directions for developing efficient UAV-ISAC systems. | *** | *** | *** | *** | *** | *** |

× *Not Covered, *Preliminary Level, ** Partially Covered, *** Fully Covered*

accuracy. A pivotal factor in effective trajectory optimization is the thorough mapping of user distribution. However, striking a balance between S&C while addressing path loss imbalances presents a considerable challenge. Moreover, the adaptability of UAV trajectories in dynamic environments is vital for improving resource allocation and enhancing overall performance. By leveraging these adaptable trajectories, UAVs can better respond to changing conditions, ultimately leading to more efficient operations.

- **Future Research Directions:** The survey highlights several critical challenges facing UAV-based ISAC systems. These challenges encompass the development of efficient algorithms, the formulation of energy-efficient mobility strategies, ensuring secure communications, addressing environmental impacts, and establishing standardized protocols.

In this survey, the layout is as follows: Section II introduces UAVs, ISAC, and their wireless network integration, detailing UAV functionalities as BSs, relay nodes, and sensing platforms. The section further explores UAV-based ISAC systems, highlighting their advantages and the challenges they face. Section III reviews practical UAV-based ISAC applications, emphasizing channel estimation (CE), target/beam tracking, throughput maximization, WSR and sensing balance, delay and AOI reduction, EE enhancement, and security improvements. Each subsection features summary tables outlining UAV roles, sensing types and metrics, methodologies, operational variables, objectives, and strategy comparisons. Section IV reflects on lessons learned, identifies ongoing challenges, and proposes future research directions. Finally, Section V concludes the survey on UAV-based ISAC systems within the context of evolving wireless communications. Our survey taxonomy is presented in Fig. 2.

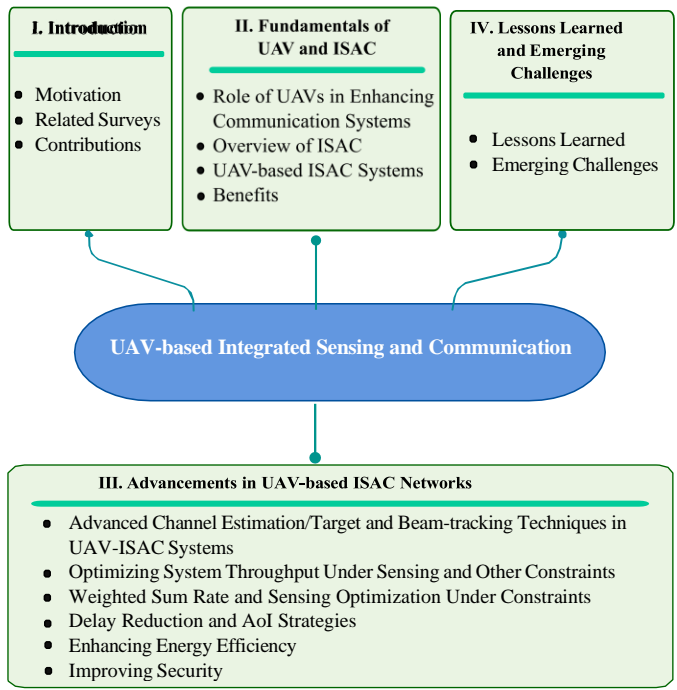

Fig. 2: Taxonomy of the paper.

## II. FUNDAMENTALS OF UAV AND ISAC

This section explains how UAVs improve communication systems as Aerial BS, Aerial Relays, and Aerial Transceivers, enhancing flexibility and efficiency. It introduces ISAC, which integrates S&C functions to optimize network performance. The discussion includes key Types of Sensing—Monostatic, Bistatic, and Multistatic—critical for UAV-based systems. Lastly, it emphasizes the advantages of UAV-based ISAC systems.

### A. Roles of UAVs in Enhancing Communication Systems

UAVs, or drones, are versatile platforms capable of autonomous or remote operation [34]. Their key attributes—ease of deployment, high mobility, low maintenance costs, and hovering ability—make them ideal for diverse civilian applications [13]. These applications include real-time traffic management, remote sensing, wireless coverage, goods delivery, search and rescue, precision agriculture, security, surveillance, and infrastructure inspection [35]. UAVs are categorized by function, size, flight altitude, and endurance [36]. Originally utilized in military contexts for remote surveillance and combat to protect pilots, UAVs now fulfill diverse civilian roles, including aerial inspections, traffic management, disaster response, delivery services, precision agriculture, and telecommunications [37]. These roles include aerial inspections, surveillance, precision agriculture, traffic management, disaster response, delivery services, and telecommunications [37].

*1) UAV as Aerial BS:* UAVs function as aerial BSs, offering wireless coverage in areas lacking traditional ground-based BSs. They facilitate swift network connectivity, particularly in emergencies and large events. This setup ensures low latency, enhanced reliability, and optimal power consumption. Moreover, they leverage advanced technologies like mmwave, free space optical communication (FSO), and LiDAR, which are essential for extending coverage in densely populated regions or disaster-stricken areas [38]–[40]. These technologies play a crucial role in expanding coverage in densely populated areas or regions affected by disasters [38]–[40].

*2) UAV as Aerial Relay:* UAVs serve as mobile relays, enhancing network coverage by strengthening communication links between base stations (BSs) and user equipment, particularly when direct connections are obstructed. Their dynamic repositioning capability improves network performance and coverage in real time. This adaptability is especially beneficial in challenging environments like urban areas or remote locations [41]–[43].

*3) UAV as Aerial Transceiver:* UAVs serve as transceivers or aerial user equipment (UE), forming links with terrestrial and aerial networks. By sharing essential operational data, including altitude and flight mode, UAVs significantly contribute to improving network performance, especially in the face of challenges such as latency and throughput. Due to their line-of-sight connections, UAVs face different interference scenarios that vary from those encountered by ground-based devices. As a result, this operational landscape requires advanced control and security protocols to maintain stable connectivity [44]–[46].

In these capacities, UAVs play a crucial role in improving communication systems by providing effective and adaptable solutions. They expand coverage areas, facilitate backhaul operations, and ensure real-time connectivity during routine and emergency situations.

### B. Overview of ISAC

The integration of communication technologies, such as wireless networks and signal processing, with sensing technologies like radar, vision, and LiDAR into a unified framework is known as ISAC. This integration significantly boosts UAV operational efficiency by enabling multimodal synchronized data collection and communication [7], [11]. This convergence allows the development of systems that leverage shared resources, including frequency bands, hardware, and computational power [6], [18]. Sensing involves using radio signals to detect and analyze various attributes of target objects in the environment. When incorporated into a communications network, sensing transforms it into an operational "radar" sensor. Using its own radio signals, the network not only provides communication but also becomes capable of perceiving and interpreting its surroundings. This integration enables comprehensive data collection, including range, velocity, position, orientation, size, shape, imagery, and material composition of objects and devices, enhancing its ability to interact intelligently with the environment [11].

Sensing for communication uses sensing data, such as CSI and user location, to improve communication performance by optimizing processes such as channel estimation and beamforming [7], [18]. This ensures reliable communication in dynamic environments, where mobility and obstacles frequently alter network conditions. S4C's ability to adapt to these changes makes it ideal for critical applications like beam tracking, interference management, and real-time communication in urban settings and smart cities, where responsiveness is essential for maintaining seamless connectivity [3], [7]. Communication for sensing relies on communication infrastructure to support and enhance sensing tasks, such as object detection and localization. Feedback signals and shared data enable improved sensing accuracy and robustness, even in challenging conditions such as disaster-stricken areas or dense urban environments. C4S ensures effective coordination between sensing nodes, reducing redundancy and enabling efficient resource use. It is particularly valuable in applications such as surveillance, emergency response, and military operations, where accurate and timely data collection is critical [7], [9].

*1) Types of Sensing:* In ISAC systems, three primary sensing methods are used: monostatic, bistatic, and multistatic sensing [9], [27]. Each method has its own strengths and is suited for different types of applications.

*a) Monostatic Sensing:* Monostatic sensing involves a single unit that transmits and receives the signal. This method is commonly used in radar systems for tasks like Earth observation and environmental monitoring [27]. The simplicity of using one unit for both transmitting and receiving makes it efficient, particularly when the size or power of the system is limited. By analyzing the signal's time delay and Doppler shift, it can determine key parameters such as distance and speed.

*b) Bistatic Sensing:* Bistatic sensing uses two separate units: one for transmission and another for reception. This separation allows for more flexible sensing geometries and better coverage by offering multiple angles of observation. It is particularly effective for tracking moving objects or mapping terrain, as it reduces interference and improves detection in complex environments [18].

*c) Multistatic Sensing:* Multistatic sensing expands on bistatic sensing by using multiple transmitters and receivers at various locations. This configuration improves spatial coverage, improves detection accuracy, and provides more comprehensive monitoring from different perspectives. Multistatic sensing is ideal for large-scale applications like surveillance or environmental monitoring, where tracking multiple targets or ensuring high reliability is crucial [7], [9].

### C. UAV-based ISAC Systems

UAVs are a versatile platform for modern communication and sensing systems, thanks to their flexibility, mobility, and cost-effectiveness. ISAC combines S&C functions into a single system, optimizing both through shared resources. When applied to UAVs, ISAC enhances operational capabilities by integrating environmental monitoring and object detection with communication technologies. This synergy allows UAVs to transmit data in real-time while maintaining heightened environmental awareness, both crucial for mission success.

Traditionally, wireless communication and radar sensing have been separate systems, each occupying distinct parts of the spectrum. However, with the expansion of cellular networks into higher frequency ranges—sub-6 GHz, mmWave, and THz—and the increasing demand for multifunctional networks, ISAC has garnered growing interest [9], [10]. By leveraging shared spectrum for both S&C, ISAC addresses spectrum scarcity and enables the use of shared RF infrastructure, resulting in more compact, lightweight, energy-efficient and cost-effective systems. This dual capability is particularly beneficial for UAV networks, where size, weight, and power (SWaP) constraints are critical, and radar sensing plays a pivotal role in integrating UAVs into B5G and 6G networks. ISAC fundamentally allows communication and radar functions to operate as distinct systems with separate waveforms, employing careful interference management through resource allocation [47]. For example, a UAV transmitting radar waveforms for sense and avoid (SAA) must mitigate interference with communication links, such as those between BSs and ground users.

A more advanced approach is dual function radar-communications (DFRC) [48], integrating radar and communication into a single system. DFRC is particularly suitable for UAV networks, allowing systems like BSs to simultaneously communicate with ground users while monitoring both ground and airspace for adversaries. DFRC can be categorized into radar-centric, communication-centric, or integrated designs. In radar-centric DFRC, radar-oriented waveforms carry communication data, but with limited data rates [49]. In contrast, communication-centric DFRC systems prioritize communication, with radar functionality as an add-on. These systems often employ waveforms like orthogonal frequency division multiplexing (OFDM), optimizing the balance between communication and radar performance [49], [50] However, in high-mobility environments such as UAV networks, OFDM performance may degrade due to sub-carrier misalignment [51]. Integrated radar-communication designs, free from radar- or communication-centric constraints, enable further optimization. For instance, a hybrid approach combining radar and communication waveforms with optimized transmit covariance matrices can preserve radar beam patterns while ensuring minimum SINR for communication users [52].

### D. Benefits

Integrating S&C in UAV-based ISAC systems provides key advantages. By sharing resources like power and bandwidth, ISAC-enabled UAVs can execute both tasks in a unified manner. This strategy optimizes resource utilization, leading to enhanced operational cost and efficiency [53]. For instance, using a common frequency band for both S&C minimizes energy consumption. Additionally, these systems support real-time data sharing, allowing UAVs to swiftly transmit sensor data to other UAVs or ground stations. This capability is vital for time-sensitive missions. Real-time communication and processing can be modeled with delay-tolerant communication systems to refine data transmission for both delay and reliability [54]. ISAC also enhances reliability through adaptive communication protocols. Moreover, ISAC promotes autonomous operations, allowing UAVs to modify their sensing strategies or flight paths based on communication feedback. This flexibility significantly enhances mission efficiency [55], [56]. This adaptability greatly enhances mission efficiency [55].

## III. ADVANCEMENTS IN UAV-BASED ISAC NETWORKS

This section reviews recent advancements in UAV-based ISAC systems, with a keen focus on strategies designed to optimize performance across multiple dimensions. Organized into clearly defined subsections, the discussion addresses unique challenges alongside various optimization techniques, aiming to enhance the functionality and efficiency of UAV-ISAC systems. The insights highlight the potential of these technologies to improve communication reliability, enhance sensing accuracy, boost system throughput, increase EE, reduce delay AoI, and strengthen security measures.

Subsection III-A explores advanced CE and target-beam tracking methods in UAV-ISAC systems. These techniques enable UAVs to maintain optimal communication links while precisely tracking targets, which enhances operational precision and reliability. In Subsection III-B, we examine strategies to augment system throughput amid sensing constraints. This section analyzes the balance between high communication throughput and limitations from concurrent sensing tasks. By employing cutting-edge optimization strategies, we can enhance system performance while adhering to operational limits. Subsection III-C focuses on dual optimization of communication and sensing capabilities through WSR and sensing optimization under specific constraints. It evaluates methodologies to maximize WSR while considering demands related to sensing and system limitations. This dual optimization ensures the effective operation of both communication and sensing functions, preserving their individual efficiencies. In Subsection III-D, we address strategies to minimize delays in UAV-ISAC systems. Timely data transmission is crucial for real-time applications like autonomous navigation and surveillance. This section emphasizes techniques to reduce delay and latency, ensuring swift communication and sensing for real-time decision-making. Subsection III-E shifts focus toward enhancing EE within UAV-ISAC systems. Given UAVs' energy constraints, optimizing energy utilization without compromising performance is essential. This segment outlines approaches for reducing power consumption, including energy-efficient algorithms and power-aware communication protocols, extending UAV operational time while maintaining optimal functionality. Lastly, Subsection III-F emphasizes the importance of bolstering security in UAV-ISAC systems. Security is critical as these systems face threats like eavesdropping, jamming, and unauthorized access. This section evaluates advanced security measures to protect both communication and sensing functions, ensuring data confidentiality within UAV networks. Collectively, these subsections highlight the complexity of UAV-ISAC systems and ongoing advancements to address their challenges. From optimizing communication and sensing to enhancing EE and security, these sophisticated techniques are vital for advancing next-generation UAV-based applications.

### A. Advanced Channel Estimation/Target and Beam-tracking Techniques

This subsection examines three interconnected techniques essential for optimizing UAV-enabled ISAC systems: CE, target tracking, and beam tracking. In CE, advanced methods enhance communication reliability by accurately modeling UAV movement and environmental conditions, with hybrid systems like RIS further improving estimation accuracy and system efficiency. Target tracking techniques, particularly extended Kalman filters (EKF), optimize UAV trajectories for precise target tracking while addressing key communication and operational constraints. Beam tracking methods, including dual identity association (DIA) and EKF, enhance beam alignment and tracking, ensuring stable communication in dynamic environments. Beam tracking methods, including dual identity association (DIA) and EKF, improve beam alignment and tracking, ensuring stable communication in dynamic environments.

This study [57] examines UAV-enabled networks for tracking ground users, employing ISAC technology for time-delay and Doppler measurements alongside communication tasks. To enhance tracking accuracy, the authors propose an EKF-based framework that utilizes the temporal correlation of user locations. The system also aggregates multiple measurements to estimate user velocity, effectively mitigating high error rates common in single BS-based velocity estimation. Results demonstrate that UAV-based ISAC considerably surpasses traditional single-BS methods regarding tracking accuracy.

The other paper [58] examines hybrid reconfigurable intelligent surfaces (HRIS) that seamlessly blend programmable reflections with sensing capabilities. This integration positions HRIS as vital for the progression of future wireless networks and ISAC applications. The authors emphasize UAV networks enhanced by HRIS, meticulously adjusting key parameters like phase profile, reception combining, and power allocation. This strategic design enables the joint estimation of UAV-HRIS and HRIS-base-station channels, as well as the angle of arrival (AoA) for the LoS component in the UAV-HRIS channel. Additionally, the study computes the Cramér-Rao lower bounds for CE and evaluates their approach's efficacy using CE error and LoS AoA accuracy metrics. Simulation results demonstrate that the proposed methodology markedly enhances the performance of HRIS-enabled ground-to-UAV communication systems.

The authors in [59] investigate ISAC technology, emphasizing its vital role in enhancing UAV communications. They design ISAC systems tailored for UAV-assisted communication, addressing UAV jitter effects. Their methodology involves developing a spatial model that captures UAV movement and attitude fluctuations impacting the communication channel. To tackle these challenges, they propose an efficient CE technique that utilizes ISAC to simultaneously analyze UAV sensing, communication, and control processes, optimizing overall system performance. Extensive simulations validate the effectiveness of their approach, demonstrating its potential to significantly improve UAV communication systems.

This paper [60] delves into the increasing significance of ISAC technology, particularly regarding UAVs, and underscores its transformative potential for the future of wireless communication systems. ISAC has emerged as an effective solution for navigating the challenges presented by LoS blockages frequently encountered in ground communications, especially within the millimeter-wave and terahertz frequency bands. However, recent research reveals that factors such as variable wind speeds and vibrations induced by UAV wings can introduce jitter, adversely affecting communication performance. As the development of ISAC technology progresses, it becomes crucial to assess the influence of UAV jitter on sensing capabilities. To address this concern, the authors present a UAV jittering model and investigate its implications on target angle sensing through a time-division wave scanning approach. The findings indicate that UAV jittering significantly undermines the accuracy and resolution of target sensing, a conclusion that is further supported by simulation results.

TABLE III: Summary of Channel Estimation/Target and Beamtracking Techniques in UAV-ISAC Systems

| Ref. | Year | UAV Details | | Communication and Sensing Details | | | CSI | Methodology | Opt. variables | Objective |
|---|---|---|---|---|---|---|---|---|---|---|
| | | Role | # (UAVs) | #(UEs) | Sensing Type | Sensing Metric | | | | |
| [57] | 2022 | T&S | M | S | GU tracking | Target position estimation | Avail. | Multiple drone measurements, EKF | – | To minimize position and velocity estimation errors. |
| [58] | 2022 | T | S | GBS to UAV via HRIS | UAV direction sensing relative to HRIS | Channel estimate MSE | Avail. | root MUSIC | UAV-HRIS and HRIS-BS channels | To maximize UAV-HRIS and HRIS-BS channel estimation accuracy. |
| [59] | 2023 | T&S | S | Only GBS to UAV commun. | GPS, IMU | Channel estimation | Avail. | UKF-based fusion | BF, trajectory | To minimize channel estimation error. |
| [60] | 2023 | T&S, BS | S | M | Target angle sensing | – | Avail. | Time-division beam searching | – | Study the impact of UAV jittering on sensing performance |
| [61] | 2023 | Receiver | M | Only GBS to UAV commun. | UAV tracking | Beam association error | Avail. | EKF, VBO | BF | To maximize the achievable rate. |
| [62] | 2023 | T&S | S | S | AoA / AoD sensing | MSE of AoA / AoD estimation | Avail. | Phase / amplitude compensation | Elevation and azimuth angle from UAV-BS to the terrestrial user | To maximize the system SNR. |
| [63] | 2023 | T&S | M | S | Accurate GU tracking | position estimation CRLB | Avail. | EKF for GU tracking, SCA for trajectory optimization | Trajectory | To maximize downlink communication rate. |
| [64] | 2024 | T&S | S | Only GBS to UAV commun. | GPS, IMU | position / orientation estimation accuracy | Avail. | EKF-based data fusion | BF | To maximize downlink SE |
| [65] | 2023 | T&S | M | M | Target detection | Target detection probability | Avail. | Signal reuse for detection, SCA for UAV location optimization | UAV location and TX power | To maximize the minimum target detection probability. |
| [66] | 2024 | T&S | M | IoT devices, GBS | Gather IoT status information | Radar estimation rate | Avail. | Iterative optimization algorithm | UAV task scheduling, trajectory, and Tx power | To maximize informaton rate and RER. |
| [67] | 2024 | Receiver | M | Only GBS to UAV commun. | UAV tracking | – | Avail. | EKF, MUSIC | BF | To maximize the achievable communication rate. |
| [68] | 2023 | T&S | S | M | Target detection | Min radar beam gain | Avail. | Reinforcement learning | Trajectory | To maximize the min beam gain at the desired sensing angle of target. |
| [69] | 2024 | T&S | S | M | Target detection | Weighted sum of predicted PCRB | Avail. | EKF, SCO | Trajectory | To maximize the achievable data rate. |

The challenge of beam alignment in millimeter-wave (mmWave) UAV networks is addressed through the use of ISAC technology. A novel dual identity association (DIA)-based ISAC method is introduced in [61], which distinguishes physical identities (P-IDs) from echo signals to achieve fast and accurate beamforming for multiple UAVs. By employing EKF to track and predict P-IDs, the system ensures precise beam alignment, even in dynamic network environments. Numerical simulations show that this DIA-based ISAC approach substantially improves both association accuracy and communication performance compared to conventional methods.

In UAV-enabled systems, beam tracking is a critical design element. The authors [62] propose an improved unscented Kalman filter (UKF)-based beam tracking method that utilizes ISAC to estimate the angle of arrival (AoA) and angle of departure (AoD) of the echo signals. The method accommodates UAV mobility and addresses estimation errors associated with conventional UKF-based approaches. Simulation results demonstrate that the proposed method achieves nearly the same SNR performance as baseline methods assuming perfect AoA/AoD estimation while providing accurate tracking in dynamic UAV networks.

This paper [63] explores a UAV-enabled ISAC system, where multiple UAVs operate concurrently to monitor a moving ground user (GU) while also facilitating downlink communication. The central focus of the study is the optimization of UAV trajectories to enhance both S&C performance. To achieve this, the EKF is employed to predict and track the GU's movements, using range measurements derived from BS sensing echoes. A carefully formulated weighted optimization problem guides the design of UAV trajectories and manage the GU-UAV associations, while factoring in real-time communication rates and Cramér–Rao bounds (CRB) for tracking accuracy. Furthermore, the framework considers various additional constraints, such as power consumption, travel distance, and collision avoidance, which are critical for practical implementations. To address the nonconvex characteristics of this optimization challenge, the authors introduce an iterative algorithm based on successive convex approximation (SCA). The results from simulations indicate that the proposed algorithm successfully tracks the GU while concurrently satisfying the communication and sensing requirements.

In a separate study [64], the focus is on advancements in massive MIMO technology, recognized as a promising solution for addressing interference issues in cellular-connected UAV communications. The authors present an innovative technique that combines wireless and sensor data to refine beam alignment within UAV MIMO environments. Their approach introduces a predictive beamforming framework, which encompasses both the frame structure and a predictive beamformer. This framework employs an EKF to effectively merge channel and sensor information. Simulation results demonstrate that this combined approach substantially improves the accuracy of position and orientation estimates, thereby enhancing SE in cellular UAV communications.

The authors in [65] explore a system consisting of multiple UAVs, each responsible for transmitting communication signals to designated GUs while concurrently detecting ground targets using the same signals. Their primary objective is to maximize the minimum detection probability across the target area. To achieve this goal, they focus on the joint optimization of UAV positioning and power control, while carefully adhering to several constraints. These constraints include maintaining the signal-to-interference-plus-noise ratio (SINR) for the GUs, complying with the maximum transmit power limits for each UAV, and ensuring a minimum distance between the UAVs to avert potential collisions. In the specific case involving two UAVs, the authors successfully derive a closed-form optimal solution. For systems that involve multiple UAVs, an iterative algorithm based on alternating optimization (AO) and SCA to find an effective solution. Numerical results indicate that the proposed approach significantly outperforms other benchmark schemes in terms of detection performance.

In [66], the exploration of UAVs as multifunctional platforms within the IoT focuses on their capacity to deliver both S&C services to IoT nodes, particularly in emergency scenarios where BSs are absent. The proposed UAV-assisted ISAC system assesses sensing performance through the radar estimation rate, a metric rooted in information theory. The authors present a coordinated optimization approach that recalibrates UAV task scheduling, power allocation, and three-dimensional flight parameters, aiming to enhance the radar estimation rate while ensuring that the communication rate remains consistent. To address this optimization challenge, the authors devise a three-layer iterative algorithm. Simulation results indicate that optimizing the flight parameters of UAVs can lead to a marked improvement in sensing performance.

A new ISAC-assisted beam tracking solution is presented in [67] for UAV-enabled systems operating in multipath channels, where reflected echoes are used to measure kinematic parameters. EKF is applied to predict angles, and a refined beam tracking method is employed based on EKF results. The proposed solution demonstrates significant advantages over traditional feedback-based techniques, effectively closing the gap between observed radar angles and optimal alignment in multipath conditions.

The paper [68] also investigates an ISAC system supported by RIS for UAVs. The focus is on optimizing UAV trajectories to maximize beam gain at a target sensing angle while satisfying SNR and communication constraints with ground users. The problem is addressed using an improved reinforcement learning algorithm to optimize UAV trajectories. Simulations demonstrate that the deployment of IRS in UAV-based ISAC systems enhances both communication and sensing capabilities.

In the context of target tracking, the authors [69] propose a UAV-enabled ISAC system that focuses on both position and velocity estimation, as accurate velocity estimation is critical for maintaining communication links and ensuring real-time responses. Using EKF, the proposed system minimizes the predicted posterior Cramér-Rao bound (PCRB) for relative position and velocity estimation by optimizing the UAV's trajectory. The SCA is employed to derive an efficient solution, with numerical results demonstrating that the method achieves optimal performance under specific conditions by maintaining a fixed elevation angle and zero relative velocity. The results further highlight interesting trade-offs in system performance resulting from the fixed elevation angle.

**Summary:** This review examines advances in UAV-enabled ISAC systems, critically highlighting the contributions and relevant technical limitations of each study. In [60], the impact of UAV jittering on S&C is explored, with the authors proposing a model to quantify its effect on target sensing accuracy. Although insightful, the study lacks a focus on real-time adaptive compensation methods to counter jittering, limiting its practical applicability. Similarly, [64] enhances the alignment of MIMO beams by integrating wireless and sensor data through a predictive beamforming framework using EKF. However, it does not address potential latency issues or inaccuracies arising from rapid environmental changes, which could degrade performance in dynamic settings. The work in [59] introduces a spatial model for UAV-aided ISAC systems, addressing UAV movement and attitude variations through joint S&C processes. However, the method assumes ideal UAV control, which limits its applicability in unpredictable multi-UAV environments. In [58], HRIS-based designs sig-nificantly improve UAV CE and sensing, but the study overlooks practical deployment issues such as the power and cost constraints associated with large-scale HRIS implementation. The trajectory optimization framework in [63] improves multi-UAV tracking and communication, but the iterative algorithm is computationally expensive and less scalable for dense UAV networks. In [65], joint UAV positioning and power control optimize communication and sensing while maintaining SINR thresholds. However, the study does not consider inter-UAV interference, which can become a significant issue in dense networks. Similarly, [57] utilizes EKF for user tracking in UAV-enabled ISAC systems but assumes perfect measurement accuracy, which is challenging in high-noise environments or with limited sensing resources. The work in [66] focuses on optimizing UAV task scheduling for IoT applications, enhancing radar estimation rates. However, it assumes static IoT nodes, limiting the model's adaptability to scenarios involving mobile nodes or dynamic task priorities.

The DIA framework in [61] improves beam alignment but lacks a discussion on scalability and inter-UAV coordination, which are critical for large-scale deployments. Similarly, [67] develops a beam tracking method based on EKF for multi-path channels, but its computational complexity could pose challenges in real-time multi-UAV systems. The UKF-based method proposed in [62] improves the estimation of AoA and AoD, but assumes ideal UAV mobility predictions, reducing its applicability in unpredictable environments. The reinforcement learning-based approach in [68] optimizes IRS-enabled UAV trajectories but requires extensive training data and does not generalize well to different scenarios or UAV configurations. Finally, [69] focuses on optimizing the trajectory for the estimation of position and velocity, achieving high accuracy using SCA. However, the approach assumes fixed elevation angles and limited environmental complexity, making it less suitable for dynamic or cluttered real-world settings. Each of these studies presents valuable advancements, yet they face challenges such as scalability, computational overhead, environmental adaptability, and practical deployment feasibility. Addressing these limitations through robust, real-world-validated models and scalable frameworks will be crucial for advancing UAV-enabled ISAC systems.

*B. Optimizing System Throughput Under Sensing and Other Constraints*

This subsection outlines strategies to optimize throughput in UAV-based ISAC systems managing SC, as shown in Fig. 3. Balancing functions under spectrum, power, and mobility constraints is challenging. Time-frequency resource allocation segments the integrated signal into radar (R) and radar communication (RC) modes to optimize spectrum usage. Optimization algorithms, such as AO and deep reinforcement learning (DRL) facilitate effective resource distribution. In emergencies, UAVs function as relay nodes or BSs, employing rate-splitting multiple access (RSMA) for optimal communication during disaster recovery. Innovations like reconfigurable intelligent surfaces (RIS) and movable antenna arrays (MA) improve signal quality and coverage, increasing throughput in dynamic environments. These advancements in resource allocation and modulation significantly enhance the throughput of UAV-based ISAC systems, enabling efficient dual-function operation.

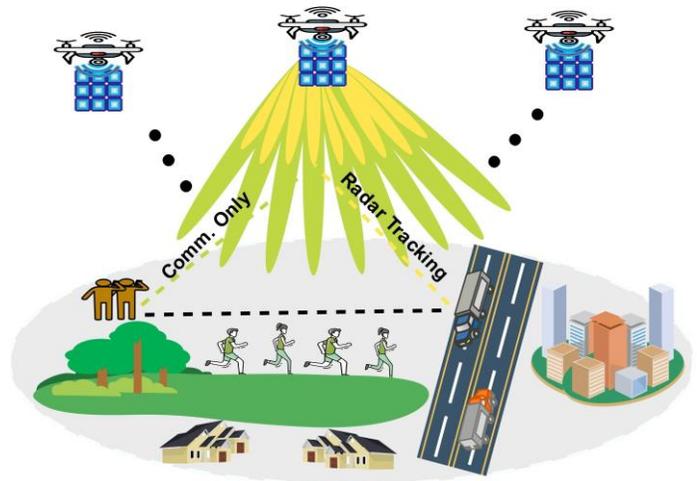

Fig. 3: An illustration of ISAC system for throughput maximization

This paper [70] presents a novel multiple access method for ISAC-enabled UAV ad hoc networks, enabling simultaneous S&C. The approach employs an integrated signal and introduces a spatial division method utilizing a multi-beam framework with tunable analog antenna arrays. A time-frequency resource allocation scheme is developed, separating the integrated signal into radar and communication modes. In addition, a new channel assignment procedure is introduced to optimize spectrum usage based on packet arrival rates. The medium access method's performance is analyzed using a Markov model. Simulation results indicate that the proposed method improves UAV node throughput by incorporating sensing information. Another paper [71] presents a UAV-assisted ISAC system that integrates radar and wireless communication for simultaneous sensing and user communication. The UAV utilizes radar data and channel state information to predict potential blockages in the radio propagation environment. A joint design methodology optimizes transmit beamforming vectors for the UAV and radar signals, enhancing communication rates. A trade-off factor is introduced to balance S&C performance. Simulation results reveal that the UAV's blockage prediction significantly boosts ISAC system performance.

This paper [72] presents a UAV-assisted full-duplex (IBFD) in-band OTFS-ISAC system for V2X communications in the 2 frequency band, employing subarray-based hybrid beamforming and strong air-ground channel line of sight. Vehicle motion parameters are tracked using a maximum log-likelihood (ML) estimator. Simulations indicate that the ML estimator performs similarly for vehicles in both OFDM and OTFS-based IBFD-ISAC systems, with OTFS enhancing SE by minimizing cyclic prefix overhead. Another paper [73] introduces a mobility-aware resource allocation approach for joint S&C. Findings indicate that the proposed algorithm enhances resource distribution between S&C in ISAC-enabled UAV networks. Similarly, this paper [74] presents a novel ISAC waveform design for mmWave UAV communications through OCDM, utilizing orthogonal chirp signals for dual S&C capabilities. The holistic design integrates OCDM with advanced FMCW, dedicating one subcarrier for sensing while enhancing communication data rates through others. This OCDM-FMCW approach reduces hardware complexity in analog-to-digital converters, offering an energy-efficient ISAC solution for resource-limited UAVs. Simulation results show this method outperforms traditional techniques like OFDM and OTFS, optimizing S&C performance and hardware complexity. This OCDM-FMCW approach significantly reduces hardware complexity, especially in the analog-to-digital converter, providing an energy-efficient ISAC solution suitable for resource-limited UAVs. The simulation results demonstrate that the proposed method is effective and superior, outperforming traditional techniques such as OFDM and OTFS by optimizing the trade-offs between S&C performance, as well as hardware complexity.

The authors [75] examine an ISAC downlink system using a UAV relay that decodes and forwards data, ensuring blockage-free communication. Their objective is to maximize the SR in the UAV-assisted ISAC system while ensuring positioning accuracy through optimized transmit power allocation. In a dynamic 3D environment with moving targets, optimization becomes challenging due to state transitions. To tackle this, they develop an approximated state transition model for 3D objects and formulate a strictly convex optimization problem, resulting in a unique power allocation solution. Numerical results show that their proposed power allocation method surpasses the baseline feedback-based beam training approach in achievable SR. In paper [76] the authors present an adaptable ISAC (AISAC) mechanism in a UAV-based system, allowing the UAV to perform on-demand sensing during communication. Sensing duration varies according to application needs, rather than aligning with communication duration, preventing unnecessary sensing and optimizing radio resource use, thus enhancing overall performance. The aim is to maximize average system throughput by jointly optimizing communication and sensing beamforming, as well as the UAV's trajectory, while adhering to quality-of-service standards for both S&C. An AO algorithm is proposed to solve the non-convex optimization problem, iteratively refining beamforming and UAV trajectory for a suboptimal solution. Numerical results validate the mechanism's effectiveness and the algorithm's efficiency. Numerical results confirm the effectiveness of the proposed mechanism and the algorithm's efficiency.

This paper [77] examines a RIS-enabled ISAC system, utilizing a UAV-mounted RIS to improve signal quality and enhance connectivity through mobility and the adjustment of signal phase and amplitude. A max-min communication rate problem is presented, focusing on the joint optimization of the beamforming vector, RIS phase shift, and UAV trajectory, while meeting power consumption and sensing constraints. The problem is non-convex and challenging to solve optimally due to variable interdependence. To address this, it is reframed as a sequential decision-making task, and a DRL-based solution is proposed for efficient resolution. Numerical results indicate that the proposed algorithm outperforms existing benchmarks. [78] presents a coordinated RSMA-based ISAC (CoRSMA-ISAC) approach for emergency UAV systems in search and rescue operations. In this system, multiple UAVs communicate with several communication survivors (CSs) and detect potentially trapped survivors (TS). The authors formulate an optimization problem to maximize the WSR while meeting sensing SNR requirements. To tackle this non-convex problem, the study decomposes it into three subproblems: UAV-CS association, UAV deployment, and beamforming optimization with rate allocation. An iterative approach utilizing K-means clustering, SCA, and semidefinite relaxation (SDR) efficiently solves the subproblems. Simulation results indicate that the CoRSMA-ISAC scheme outperforms traditional methods such as SDMA, NOMA, and OMA in communication and sensing performance.

This paper [79] examines a resource allocation problem in a multi-UAV-assisted ISAC system with dual-functional UAVs for radar sensing and communication with ground users. It optimizes UAV trajectories, user associations, and beamforming to maximize the sum-weighted bit rate for users while ensuring detection beampattern gain for the target. To tackle the mixed-integer non-convex optimization challenge, the authors decompose it into two subproblems using an AO framework. For user association and beamforming design, a novel algorithm utilizes matching theory (MT) and fractional programming (FP). Sequential quadratic programming addresses the non-convex UAV trajectory subproblem, deriving a suboptimal solution from successive quadratic programming (QP) problems. The subproblems are iteratively solved until convergence, producing a stable solution. Simulation results show that the proposed strategy significantly outperforms benchmark methods, including the deferred acceptance algorithm, K-means algorithm, and heuristic approaches, improving both sensing beampattern gain and communication rate.

Unlike the SNR sensing metric-based works [77]–[79], this paper [80] explores rates with multiple UAVs providing ISAC services to IoT nodes. It introduces radar mutual information (MI) to evaluate ISAC sensing performance from an information theory perspective. The objective is to maximize the minimum communication rate for fairness among IoT nodes, necessitating joint optimization of node scheduling, transmit power, and UAV 3D trajectories, while adhering to radar MI constraints. The complex non-convex problem is divided into three subproblems: UAV scheduling, transmit power, and 3D trajectory optimization. A three-layer iterative optimization algorithm is proposed to address these subproblems, yielding near-optimal solutions. Simulation results indicate that this optimization approach significantly improves communication rates while satisfying radar MI constraints and ensuring equitable communication for all nodes. Similarly, the authors in [81] explore a UAV-assisted ISAC network with moving ground users on constant-velocity trajectories. It presents a global optimal trajectory design scheme offering a continuous analytical solution, contrasting traditional discrete numerical methods. To tackle the challenges of maximizing performance across infinite time slots, the paper projects trajectories onto a user-relative coordinate frame, creating a location-dependent function similar to an artificial potential field (APF). This approach reformulates the optimization into a shape-determination problem for a density-varying catenary within the APF. The analysis of forces results in a second-order differential equation that describes the catenary's topology, shaped by boundary conditions like location, orientation, and turning curvature at waypoints. The continuous solution minimizes within a low-dimensional parameter space, enabling a rapid and adaptable design process. Complexity analysis reveals a significant reduction compared to traditional methods. The study affirms global optimality, existence, and uniqueness of the solution for S&C services under broad conditions. Simulations demonstrate the method's efficiency, highlighting global optimality, reduced complexity, and flexibility in managing avoidance, crossing, and G-force limits. While in [82], the authors redefine optimal UAV trajectory design as a catenary shape problem, emphasizing the minimization of potential rather than maximizing performance. They model a partial distribution of GU locations as matter with areal mass density in an artificial potential field (APF). To derive the optimal trajectory, the authors present a second-order mechanical equation that reflects the catenary's shape, grounded in its static equilibrium state when potential is minimized. This approach contrasts with traditional path discretization methods, yielding a continuous-form second-order equation that significantly compresses the path. Numerical results demonstrate that this method surpasses conventional UAV trajectory optimization, delivering optimal trajectories with low computational complexity while flexibly adapting to uncertain GU locations.

Different from the SNR sensing metric works in [80]–[82] This paper [83] focuses on beam-pattern gain, enhancing S&C capabilities through maneuverability and robust air-to-ground LoS links. The study addresses a UAV-AP with a vertically oriented ULA, integrating S&C signals for multiple users and ground target detection. Two scenarios are examined: a quasi-stationary UAV at an optimized location and a fully mobile UAV. The aim is to optimize UAV positioning and transmit beamforming to maximize weighted sum-rate throughput for users while achieving sensing beampattern gain under power and flight constraints. These challenges are nonconvex, as UAV location or trajectory influences steering vectors and interacts with transmit beamforming variables. Efficient optimization-based algorithms are developed for high-quality suboptimal solutions. Numerical results demonstrate that the proposed designs outperform benchmark schemes. Another paper [84] introduces a novel integrated periodic sensing and communication (IPSAC) mechanism, enabling a flexible trade-off between functionalities. The approach seeks to maximize system rate by jointly optimizing UAV trajectory, user association, target sensing selection, and transmit beamforming while satisfying sensing frequency and beam pattern gain criteria. The authors derive a closed-form expression for the optimal beamforming vector, reducing complexity despite the non-convex problem and integer variables. They also provide a tight lower bound for the achievable rate to assist UAV trajectory design. A two-layer penalty-based algorithm is introduced for efficient problem-solving. The study analyzes optimal achievable rate and UAV positioning with an infinite number of antennas, revealing symmetry between optimal solutions across various ISAC frames without location constraints. This insight leads to an efficient algorithm for location-constrained problems. Numerical results validate the designs' effectiveness, demonstrating a more flexible trade-off in ISAC systems compared to existing benchmarks. In [85], the authors present a novel mechanism for integrated periodic S&C within UAV-enabled ISAC systems. The goal is to maximize user rate by optimizing UAV trajectory, transmit precoder, and sensing timing while adhering to constraints on sensing frequency and beam pattern gain. The authors provide a closed-form solution for the optimal transmit precoder and achievable rate for any UAV location, aiding trajectory design despite complexity and nonconvexity. Additionally, the paper reveals structural symmetry in optimal solutions across ISAC frames without location constraints and introduces a high-quality optimization algorithm for UAV trajectory and sensing with location constraints. Simulation results confirm the effectiveness of the proposed approach, offering a more flexible trade-off in ISAC systems than traditional methods. Another paper [86] examines an ISAC system that employs a moving antenna array (MA) on a low-altitude platform (LAP) to support low-altitude economy (LAE) applications. In this scenario, a UAV serves as a UAV-enabled LAP (ULAP), performing both information transmission and sensing simultaneously for LAE tasks. To boost throughput and meet the required sensing beampattern threshold, the study formulates a data rate maximization problem, optimizing transmit information, sensing beamforming, and the positions of the MA array antennas. Given the non-convex nature of the problem and the complex interdependencies between variables, an AO algorithm is proposed, which iteratively optimizes certain variables while keeping others fixed. The numerical results indicate that the MA array-based scheme outperforms two benchmark methods in terms of achievable data rate and beamforming gain.

**Summary:** This subsection highlights advances in optimizing throughput for UAV-ISAC systems while addressing challenges in balancing S&C under various constraints. In [70], a multibeam framework with time-frequency resource allocation enhances ISAC system throughput, though scalability in dense networks remains a concern. Similarly, [71] employs joint beamforming to predict environmental blockages, but its adaptability to dynamic conditions in real-time is limited. Addressing high-mobility scenarios, [72] proposes OTFS modulation for SE but neglects inter-UAV interference. [73] further develops mobility-aware resource allocation, though mobility prediction in unpredictable environments remains a challenge.

To address hardware constraints, [74] introduces OCDM-based waveform designs for energy-efficient mmWave UAV systems, but its feasibility in multi-UAV setups is not explored. [75] ensures blockage-free communication using UAV relays, yet simplified environmental assumptions limit real-world adaptability. Dynamic adaptability is improved in [76], where an AISAC mechanism optimizes sensing durations to improve throughput, though computational complexity hinders scalability. Similarly, [77] applies RIS-assisted DRL optimization, achieving performance gains but struggling with robustness in practical scenarios.

Disaster recovery applications are explored in [78], which

TABLE IV: Summary of Throughput Maximization Schemes for UAV-ISAC Systems

| Ref. | Year | UAV Details | | Communication and Sensing Details | | | | | |
|---|---|---|---|---|---|---|---|---|---|
| | | Role | # (UAVs) | # (UEs) | Sensing Type | Sensing Metric | CSI | Methodology | Opt. Variables | Objective |
| [70] | 2022 | T&S | M | - | UAV detection | - | Avail. | Markov-chain model | ISAC channel assignment | To derive throughput perf. analysis |
| [71] | 2022 | T&S | M | M | environment/ blockage sensing | sensing BF design | Avail. | Codebook-based design | sensing and comm. BF | To maximize SR |
| [72] | 2023 | T&S | S | M | Target (vehicle) detection | ML radar tracking/ prediction | Avail. | ML estimation/tracking | BF | To maximize SE |
| [73] | 2023 | T&S | M | M | Mobile GUE detection | Mobility level detection | Avail. | Mobility-aware RA | UAV-UE association | To maximize avg. SR |
| [74] | 2024 | T&S | M | - | Multiple target sensing | Range and velocity estimation | Avail. | ISAC waveform design | - | - |
| [75] | 2023 | R&S | S | M | GUE detection | position estimation CRLB | Avail. | 3D state transition model | PA at UAV & BS | To maximize SR |
| [76] | 2024 | T&S | S | M | Target (vehicle) sensing | position estimation CRLB | Avail. | AO | BF, trajectory | To maximize avg. system throughput |
| [77] | 2024 | R&S | S | M | Multiple vehicle detection | sensing SNR | Avail. | DRL with proximal policy opt. | traj, BF, RIS PSs, info. & sensing BF, | To maximize minimum-rate |
| [78] | 2024 | T&S | M | M | Target detection | sensing SNR | Avail. | SCA, SDP | rate alloc., UAV-UE association, UAV position opt., BF | To maximize weighted SR |
| [79] | 2024 | T&S | M | M | Target (vehicle) detection | sensing SNR | Avail. | AO, FP, MT, Sequential QP | trajectory, UAV-UE association, BF | To maximize SR |
| [80] | 2023 | T&S | M | M | IoT-nodes sensing | sensing rate | Avail. | AO | trajectory, node scheduling, PA | To maximize minimum-rate |
| [81] | 2023 | T&S | S | S | Mobile GUE detection | sensing rate | Avail. | deriving differential equations | trajectory | To design trajectory |
| [82] | 2024 | T&S | S | S | GUE position sensing | sensing rate | Avail. | SCA | trajectory | To maximize rate |
| [83] | 2022 | T&S | S | M | Multiple target sensing | sensing beam-pattern gain | Avail. | AO, SCA | trajectory, BF | To maximize weighted SR |
| [84] | 2023 | T&S | S | M | Multiple target sensing | sensing beam-pattern gain | Avail. | Penalty-based opt. | sensing time selection, trajectory, BF | To maximize minimum-rate |
| [85] | 2024 | T&S | S | M | Single target sensing | sensing beam-pattern gain | Avail. | SDP | sensing time selection, trajectory, BF | To maximize rate |
| [86] | 2024 | T&S | S | M | GUEs detection | Sensing beam-pattern gain | Avail. | AO, PSO | info. & sensing BF, MA array position opt. | To maximize SR |

*S- Single, M- Multiple, H- Hybrid, R& S- Relay and Sensing, GUE- Ground UE, Comm. - Communications, max. - maximization, min. - minimize*
*BF - beamforming, RA - resource allocation, alg. - algorithm, opt. - optimization, Avail. - available, PA - power allocation, alloc. - allocation, SR - sum-rate*
*perf. - performance, info. - information, PSs - phase shifts, T& S- Transmit and Sensing, SE - spectral efficiency, ML - maximum, likelihood*

employs RSMA to optimize UAV associations and beamforming but overlooks energy constraints. [79] enhances multi-UAV ISAC systems with resource allocation frameworks, though iterative optimization poses scalability issues. In [80], fairness in resource allocation is achieved for IoT nodes, but scalability challenges persist. [81] and [82] focus on efficient trajectory designs for moving users and uncertain environments, though static assumptions reduce flexibility. Comprehensive trade-offs are addressed in [83] with dual-functional UAV designs and beamforming, though heuristic methods limit adaptability. Periodic sensing is optimized in [84] and [85], offering flexible trade-offs but requiring idealized conditions. Lastly, [86] explores movable antenna arrays for enhanced data rates but introduces hardware complexities. These studies demonstrate significant progress in the optimization of the UAV-ISAC system through innovative frameworks, mobility-aware designs, and advanced modulation techniques. However, challenges such as scalability, computational efficiency, energy management, and adaptability to dynamic environments remain critical areas for future research.

### C. Weighted Sum Rate and Sensing Optimization Under Constraints

This subsection addresses optimization techniques for WSR and sensing in UAV-enabled ISAC systems, focusing on balancing communication and sensing performance within constraints like energy limitations and trajectory design. An illustration of a UAV-based ISAC system is shown in Fig. 4, where UAVs simultaneously communicate with UEs and detect targets. A single UAV can be equipped with both transmitter and receiver antennas to transmit joint communication and sensing signals while receiving sensing echoes from the targets. Alternatively, a separate UAV can be used exclusively for receiving sensing echoes (right Rx UAV). Research targets multi-objective optimization problems, including UAV trajectory planning, power allocation, radar waveforms, and scheduling. The aim is to maximize communication rates and sensing accuracy while managing trade-offs. Methods such as AO and iterative algorithms are employed to achieve balanced solutions under energy constraints. Additionally, techniques like RIS phase shift optimization and deep reinforcement learning are utilized to enhance performance in content delivery and target sensing.

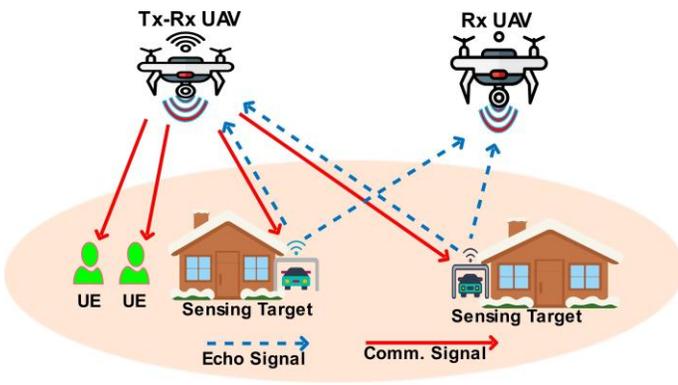

Fig. 4: An illustration of a UAV-based ISAC system where UAVs simultaneously communicate with the UEs and sense the targets. The same UAV can be equipped with both transmitter and receiver antennas to transmit joint communication and sensing signals, and to receive sensing echoes from the targets (left Tx-Rx UAV). Alternatively, the receiver UAV could be a separate UAV dedicated to receiving the sensing echoes (right Rx UAV).

This work [87] explores a rotary-wing UAV that transmits signals to a communication user while receiving echoes for target estimation. We propose a multi-stage trajectory design to optimize S&C performance, framed as a weighted optimization problem and solved with an iterative algorithm. Numerical results demonstrate the trade-off between S&C performance.

This paper [88] investigates a system where multiple UAVs serve as aerial ISAC platforms for ground users' S&C. A joint optimization problem is formulated to enhance S&C performance, focusing on user association, UAV trajectory planning, and power allocation to maximize the minimum weighted SE among UAVs. Centralized and decentralized DRL approaches address this sequential decision-making challenge. The centralized method employs the soft actor-critic (SAC) algorithm, transforming the optimization objective via symmetric group theory. Data augmentation techniques, including random and adaptive methods, enhance the SAC replay memory buffer's performance. In the decentralized framework, the multi-agent soft actor-critic (MASAC) algorithm addresses decision-making. Results show that the adaptive SAC scheme outperforms other centralized solutions in training speed and weighted SE, while the MASAC algorithm excels in early-stage training speed. The concept of robotic aerial BSs (RABS) is anticipated to increase the flexibility of ISAC systems. This work [89] addresses spatial traffic distribution through a grid-based model, framing the RABS-assisted ISAC system as a robust optimization problem aimed at maximizing the minimum satisfaction rate under cardinality constraints. The problem is reformulated into a mixed-integer linear programming (MILP) framework and solved approximately using an iterative linear programming rounding algorithm. Numerical analysis demonstrates that the proposed approach can achieve an average 28.61% improvement in minimum SR compared to fixed small cell configurations.

This paper [90] investigates an ISAC system that integrates wireless power transfer (WPT), where UAV-based radars support energy-limited communication users while sensing. In the first phase, radars conduct environmental sensing, enabling users to harvest energy from the radar signals. In the second phase (uplink), this energy powers users' information transmission to UAVs. The radar's performance relies on transmitted signals and received filters, with the transmitted signal energy facilitating the uplink. The study formulates a multi-objective design problem to optimize UAV trajectories, radar waveforms, received filters, time scheduling, and uplink power allocation, effectively balancing radar and communication performance. This is framed as a non-convex optimization problem that addresses uncertainties in user locations. The solution employs AO, fractional programming, the S-procedure, and majorization-minimization (MM) techniques. Numerical examples validate the method's effectiveness across diverse scenarios. The authors in [91] propose a RIS-assisted UAV-enabled ISAC system where a dual-functional UAV simultaneously transmits a signal to multiple users and performs sensing operations. The objective is to maximize the weighted sum of the average BS and sensing SNR, addressing the trade-off between communication and sensing. This is achieved by optimizing the RIS phase shifts, the UAV trajectory, dual-function radar communication beamforming (DFRC-BF), and user scheduling. An iterative algorithm based on AO is introduced to find a suboptimal solution. Simulation results demonstrate the approach's effectiveness compared to baseline methods, showcasing its ability to balance the trade-off between S&C functions.

This study [92] addresses the challenges of content delivery and target sensing in ISAC networks using UAVs, enabling them to store and deliver user-requested content while simultaneously sensing targets. To improve performance in both content transmission and target sensing, a utility function is defined, framing the issue as a constrained utility maximization task focusing on UAV deployment and the design of S&C precoders. Since this problem represents a mixed-integer nonlinear programming challenge, it is decomposed into two subproblems: one for user grouping and UAV deployment, and another for designing communication and sensing precoders. An alternating iteration-based algorithm is then employed to tackle these subproblems. Specifically, a mean-shift-based strategy is first introduced to group users, followed by a UAV deployment method utilizing a SCA-based iterative approach and first-order Taylor expansion. For the communication and sensing precoder design subproblem, a two-layer penalty-based SCA algorithm is proposed. Simulation results demonstrate the effectiveness of the proposed solutions.

This paper [93] investigates ISAC in the airborne domain, with UAVs functioning as communication BSs and radar systems. UAVs transmit signals for user communication while using those signals for target localization. The focus is on enhancing S&C performance through optimized UAV trajectory design and user bandwidth allocation. Communication is measured by total transmitted data, and sensing is assessed using the CRB. A trade-off objective is formulated with normalization, framing the trajectory design problem as a weighted sum optimization to balance S&C requirements. A multi-stage trajectory design (MSTD) is proposed to improve accuracy. Given the complexity of solving the problem directly, an iterative algorithm is introduced for locally optimal UAV trajectory solutions. Numerical results demonstrate the impact of the S&C performance-energy constraint trade-off on UAV trajectories.

TABLE V: Summary of Weighted SR and Sensing Optimization Schemes for UAV-ISAC Systems

| Ref. | Year | UAV Details | | Communication and Sensing Details | | | CSI | Methodology | Opt. variables | Objective |
|---|---|---|---|---|---|---|---|---|---|---|
| | | Role | # (UAVs) | # (UEs) | Sensing Type | Sensing Metric | | | | |
| [87] | 2023 | T&S | S | S | Target sensing | Target position estimation | Avail. | Gradient descent | trajectory | To jointly maximize rate and minimize CRLB for target position |
| [88] | 2023 | T&S | M | M | GUE position sensing | sensing SE | Avail. | MASAC | user association, trajectory, PA | To maximize the minimum weighted comm. and sensing SE |
| [89] | 2023 | T&S | S | M | Multiple UEs sensing | sensing rate | Avail. | LP, Rounding alg. | UAV position, subcarrier alloc. for sensing and comm. | To maximize the minimum weighted comm. and sensing rate |
| [90] | 2023 | T&S | M | M | Target detection | sensing SNR | Avail. | AO | trajectory, PA, time scheduling, radar WPT waveform | To jointly maximize the minimum comm. rate & sensing SINR |
| [91] | 2024 | T&S | M | M | Target (vehicle) detection | sensing SNR | Avail. | AO | RIS PSs, DFRC BF, trajectory UE scheduling | To maximize weighted avg. of SR and sensing SNR |
| [92] | 2024 | T&S | M | M | Multiple target sensing | sensing power | Avail. | AO, SCA | BF, trajectory, UAV-UE association | To maximize weighted SR and radar received power |
| [93] | 2024 | T&S | S | M | Multiple target sensing | position estimation CRLB | Avail. | SCA, Gradient descent | trajectory, BW alloc. | To jointly maximize rate and minimize CRLB for target position |

**Summary:** This subsection reviews advancements in UAV-enabled ISAC systems, focusing on optimizing weighted SR and sensing performance under various constraints. The study in [90] integrates WPT into ISAC, optimizing UAV trajectory, radar waveform design, and power allocation to balance S&C. However, it assumes perfect synchronization between the S&C phases, which may not hold in dynamic environments, leading to potential performance gaps. Similarly, [91] employs RIS to balance sensing SNR and communication SR, optimizing UAV trajectories and RIS phase shifts. The reliance on ideal and complete CSI at the RIS limits its applicability in practical scenarios where CSI may be incomplete or delayed.

In [87], UAV trajectory optimization highlights the trade-off between S&C but overlooks external factors like wind or obstacles that may affect UAV stability and sensing accuracy. Similarly, [93] integrates the allocation of trajectory and bandwidth to optimize performance under energy constraints but employs simplified energy consumption models, ignoring real-world energy losses caused by environmental conditions or hardware inefficiencies. The study in [88] utilizes multi-agent DRL to optimize trajectories, power allocation, and user association, thereby improving efficiency and sensing accuracy. However, it relies on large training datasets and high computational resources, making real-time deployment challenging in constrained environments. The study in [92] addresses content delivery and sensing using iterative algorithms for UAV deployment and precoder design. Although effective, it assumes a uniform user distribution, which may not reflect real-world conditions characterized by uneven terrain or urban congestion. However, [15] employs MARL to enhance the adaptability of UAVs by optimizing bandwidth and sensing accuracy. However, its computational overhead remains high and does not fully address inter-UAV interference in dense networks.

### D. Delay and AoI Reduction

This subsection discusses strategies to reduce delays and AOI while optimizing performance in UAV-enabled ISAC systems, as illustrated in Fig. 5. ISAC technology integrates Identification Friend or Foe (IFF) to minimize sensing and communication delays. Additionally, stochastic network calculus mitigates delay violations by balancing SC. For emergency operations, a two-stage deployment framework employs dynamic particle swarm optimization (DPSO) and a convolutional neural network (CNN) to expedite UAV deployment, enhancing efficiency in critical scenarios.

This paper [94] proposes ISAC technology to enhance communication and sensing. The EKF algorithm integrates communication-based location data with sensing information, thereby improving target sensing accuracy. An ISAC-based Identification Friend or Foe (IFF) method aims to reduce communication delay. Unlike conventional IFF methods, this integrated approach allows simultaneous radar S&C interrogation, significantly cutting sensing time. Simulations show that ISAC technology improves UAV sensing performance, decreases communication delay by up to 50%, and enhances target sensing accuracy by 24.2%, while maintaining equal accuracy levels for both communication and radar sensing.

TABLE VI: Summary of Delay and AoI Strategies for UAV-ISAC Systems

| Ref. | Year | UAV Details | | Communication and Sensing Details | | | | | |
|---|---|---|---|---|---|---|---|---|---|
| | | Role | # (UAVs) | # (UEs) | Sensing Type | Sensing Metric | CSI | Methodology | Opt. variables | Objective |
| [94] | 2021 | T&S | M | NA | Sensing at ground and air | Sensing accuracy | Avail. | EKF, IFF | – | To improve sensing accuracy and reduce comm. delay for safe flight. |
| [95] | 2024 | R&S, FC | M | M | Environmental or situational data | Successful sensing prob., delay violation prob. | Avail. | PCP, SNC | PA | To minimize network-layer delay to evaluate tradeoff between sensing and comm. |
| [96] | 2024 | T&S, BS | M | M | Ground target detection | Target position estimation | Avail. | DPSO, CNN | Trajectory and user association | To maximize communication throughput. |
| [97] | 2023 | T&S | S | M | Ground target detection | Target position estimation | Avail. | Dynamic programming | Trajectory | To maximize communication AoI of UAV. |
| [98] | 2023 | T&S | S | M | Data sensing | Sensing time | Avail. | SCA | Position, Comm. time, Power | To minimize the average Peak Age-of-Information. |
| [99] | 2021 | T&S | S | M | Data sensing | Sensing time, power | Avail. | SCA | Comm. time, Power, Trajectory | To minimize the maximum average Peak Age-of-Information of targets. |

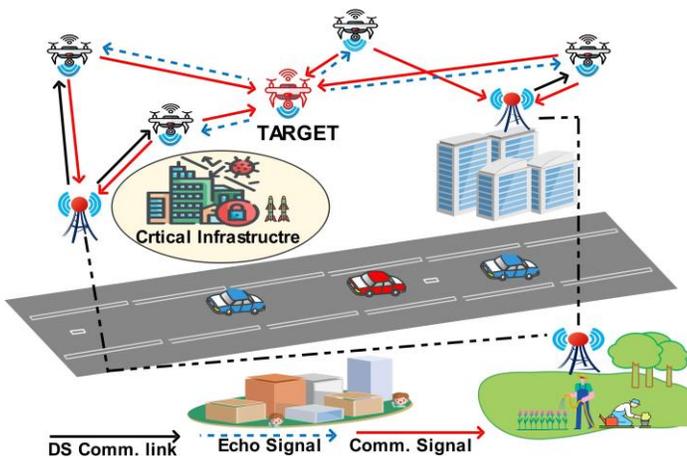

Fig. 5: An illustration of ISAC system where nodes employ integrated approach that allows S&C to occur simultaneously and in collaboration with surrounding nodes/infrastructure significantly decreasing the sensing delay.

This paper [95] addresses network-layer delay violations in an ISAC UAV network. UAVs are organized using a Poisson cluster process (PCP), enabling the calculation of successful sensing probabilities for sensory data flow. Utilizing stochastic network calculus, the paper evaluates delay violation probabilities in a two-stage sensory data transmission queue. A delay minimization problem is formulated, investigating the S&C trade-off, particularly regarding power allocation strategies. The study also analyzes delay violation probabilities during emergencies, notably when a sensing UAV encounters antenna misalignment. Through steady-state and transient analyses, the research reveals complex S&C trade-offs, offering insights into power allocation, network deployment, control module configuration, and sensory data flow management to meet performance requirements.

ISAC networks supported by UAVs are essential for emergency rescue operations after disasters, where the speed of UAV deployment directly impacts the effectiveness of the rescue. However, the complexity of deploying ISAC UAVs in emergency situations often contradicts the need for rapid deployment. To overcome this challenge, a two-stage deployment framework is proposed in this paper [96] for fast ISAC UAV deployment, consisting of offline and online stages. In the offline stage, the ISAC UAV deployment problem is formulated, with ISAC utility—combining communication rate and localization accuracy—as the objective. A dynamic particle swarm optimization (DPSO) algorithm is designed to generate an optimal UAV deployment dataset, which is then used to train a CNN. This CNN replaces the time-consuming DPSO in the online stage, enabling faster deployment decisions. In this paper [100], collaboration among multiple UAVs facilitates simultaneous multi-static radar sensing and coordinated multi-point (CoMP) transmission, thus boosting ISAC effectiveness. However, limited UAV resources complicate the optimization of dual functions. To tackle this, the paper explores an ISAC system using orthogonal frequency division multiple access (OFDMA) with UAVs, formulating a joint trajectory planning and resource allocation problem. The aim is to minimize the Cramér-Rao lower bounds (CRLB) for target localization while ensuring quality-of-service (QoS) constraints. This non-convex problem is segmented into three sub-problems, with algorithms proposed for optimal solutions. Simulations show algorithm convergence and enhanced localization performance under various communication demands compared to conventional methods. Simulations demonstrate algorithm convergence and improved localization performance across various communication demands compared to conventional methods.

The authors in the work [97] introduce ISAC technology for space-based unmanned platforms, becoming vital for future 6G wireless networks. However, in challenging environments such as remote mountain areas, the AoI in the sensing-communication services for UAVs significantly affects system performance. This paper addresses optimizing AoI through UAV trajectory planning in ISAC networks. By using the UAV's geometry as an anchor, the authors propose a dynamic programming (DP) algorithm to minimize communication AoI while ensuring precise system positioning. Simulation results show substantial improvements in S&C performance over conventional random benchmarks. Another study [98] explores UAV-assisted ISAC systems, where UAVs enhance communication and sensing by improving coverage. In this scenario, the UAV first senses a target and then sends the data to a BS for decision-making. The peak age of information (PAoI) serves as a key metric for data freshness. The authors aim to minimize average PAoI by optimizing the UAV's horizontal position, S&C times, and power allocation. For a single-target scenario, a semi-closed-form solution for optimal UAV horizontal positioning is derived using Newton's iterative method. In multi-target scenarios, the authors propose a SCA algorithm to address the nonconvex problem. Simulation results illustrate the impact of various system parameters and validate the effectiveness of the proposed optimization methods. Similarly this paper [99] investigates a UAV-assisted ISAC network where UAVs sense multiple targets using onboard radar and transmit the data to a ground BS for decision-making. The accuracy of these decisions depends on the freshness of the sensed data, quantified by PAoI. To ensure optimal PAoI performance, the authors propose a joint optimization model incorporating UAV trajectory, target scheduling, and time and power allocation for S&C. Due to the problem's non-convex nature, an efficient SCA algorithm is introduced, transforming it into a convex problem using fractional programming and penalty methods. Simulation results reveal that this approach outperforms the alternate optimization (AO) algorithm in both PAoI performance and convergence rate.

**Summary:** This subsection reviews delay and AOI reduction strategies in UAV-enabled ISAC systems, focusing on balancing S&C while addressing key performance challenges. The EKF-based ISAC approach in [94] reduces delays and improves target detection accuracy with an IFF method for simultaneous radar detection and communication. However, scalability to larger networks and potential sensor inaccuracies in complex environments remain unaddressed. In [95], network-layer delay violations in ISAC UAV systems are analyzed using stochastic network calculus, identifying power allocation strategies to minimize delays. Despite its contributions, the study simplifies environmental dynamics and lacks adaptive mechanisms for real-time changes. Similarly, [96] proposes a two-stage deployment framework that combines offline DPSO optimization with CNN-based online inference, achieving rapid UAV deployment and improved ISAC performance. However, its reliance on pre-trained models limits adaptability to unforeseen scenarios, and its effectiveness depends on the quality of the training data. Although these strategies effectively reduce delays in specific scenarios, limitations such as assumptions of ideal conditions, scalability, and adaptability to dynamic environments highlight the need for further research to ensure robust real-world ISAC performance. In [100], an OFDMA-based cooperative multi-UAV system optimizes trajectory planning and resource allocation, minimizing CRLB for localization while maintaining QoS constraints. Although effective in simulations, it assumes perfect synchronization and overlooks challenges such as inter-UAV interference and dynamic environmental factors. AoI-based approaches in [97] and [98] improve data freshness through UAV trajectory and resource optimization. However, they lack a focus on EE, critical for prolonged UAV operations, and do not account for dynamic target mobility, potentially reducing accuracy. Similarly, [99] minimizes PAoI for S&C but assumes fixed target positions, limiting its adaptability in highly dynamic scenarios.

### E. Enhancing Energy Efficiency

This subsection discusses strategies to enhance EE in UAV-enabled ISAC systems. Key approaches include optimizing radar waveforms, power allocation, and UAV trajectories for efficient S&C, alongside WPT systems that enable energy harvesting for uplink communication. MARL is employed to improve sensing performance and reduce energy consumption. Joint optimization techniques, such as SCA and fractional programming, are used to minimize power usage while maintaining QoS and sensing accuracy. Additionally, energy-efficient beam alignment and MEC integration further optimize system performance. These strategies demonstrate significant improvements in EE, sensing, and communication, highlighting the potential of UAVs to enhance ISAC systems.

This paper [101] introduces an energy-efficient computation offloading strategy for multisensor data fusion in a UAV-assisted V2X network supported by ISAC. First, a vehicle-UAV cooperative perception framework is proposed to expand the range of traffic environment awareness. Then, a computation offloading strategy is designed, taking into account offloading decisions and dynamic resource allocation. Finally, a SCA algorithm is applied to transform the nonconvex problem into a solvable convex one. The simulation results demonstrate that this approach reduces the energy consumption of the UAV and minimizes delays in processing data fusion tasks.

This paper [102] addresses the challenge of balancing high-quality communication and low-latency sensing in an ISAC system by proposing a UAV-assisted architecture that employs a flight-hover communication protocol. In this setup, the UAV communicates with the IoT devices while hovering and senses target locations during flight. To optimize the number of connected IoT devices and reduce UAV energy consumption, a DRL algorithm is developed for trajectory planning. The simulation results show that the proposed algorithm efficiently detects and collects data from the sensor devices.

In this framework [103], the radar conducts environmental sensing during the first phase (WPT), during which a UAV-based radar aids a group of energy-limited communication users while also performing sensing tasks. In this framework, the radar conducts environmental sensing during the first phase (sensing phase), allowing communication users to harvest and store energy from the radar's transmitted signals. This stored energy is subsequently used for uplink information transmission from the users to the UAV during the second phase. The radar system's performance is influenced by the transmitted signals and the receive filters, while the energy of these signals is crucial for powering the uplink communication. To address this, the study formulates a multi-objective design problem aimed at optimizing both radar and communication performance by refining the radar transmit waveform, receive filter, time scheduling, and uplink power levels. The optimization of the UAV's trajectory is also considered to ensure adequate coverage of the target area. Given the non-convex nature of the design problem, the authors present a solution based on AO, incorporating fractional programming and a majorization-minimization technique.

TABLE VII: Summary of EE Schemes for UAV-ISAC Systems

| Ref. | Year | UAV Details | | Communication and Sensing Details | | | CSI | Methodology | Opt. variables | Objective |
|---|---|---|---|---|---|---|---|---|---|---|
| | | Role | # (UAVs) | # (UEs) | Sensing Type | Sensing Metric | | | | |
| [101] | 2022 | T&S | S | S | Traffic perception environment mapping | – | Avail. | SCA | Task offloading and resource allocation for data fusion | To minimize UAV latency and energy consumption. |
| [102] | 2023 | T&S | S | S | Ground target detection | Distance between UAV and target | Avail. | DRL | Trajectory | To maximize the number of connected devices and minimize the energy consumption of the UAV. |
| [103] | 2023 | T&S | S | S | Ground target detection, surveillance | Radar SINR | Avail. | POS, ST | User scheduling, power, and trajectory | To maximize EE and sensing fairness. |
| [104] | 2023 | T&S | S | S | Ground target detection | BF accuracy | Avail. | SDR, the big-M method, and SCA | Trajectory, velocity, and BF | To minimize average power consumption, and ensure communication and sensing QoS. |
| [105] | 2023 | T&S | S | S | Ground target detection | Target position estimation | Avail. | SCA, BCD | Trajectory | To minimize UAV propulsion power. |
| [106] | 2023 | T&S | S | S | Recognize static targets | Radar estimation information rate | Avail. | BCD, ML | Sensing scheduling, number of time-slots and trajectory | To minimize UAV energy consumption and data collection time. |
| [107] | 2023 | T&S | S | S | Ground target detection | Radar mutual information | Avail. | SCA, FP | User scheduling, Tx power and trajectory | To maximize EE and min radar MI. |
| [108] | 2024 | T&S | M | M | Vehicle sensing | Sensing mutual information | Avail. | MARL | Tx power, trajectory | To improve sensing MI obtained by UAVs, improve comm. quality, save UAVs comm. energy. |
| [109] | 2024 | T&S | S | S | Ground target tracking | Target angle estimation accuracy | Avail. | EKF, BCD based FP algorithm | Number of activated antennas, BF | To maximize EE. |
| [110] | 2024 | T&S | S | S | Ground target detection, area monitoring | Ground target accumulated sensing energy | Avail. | one-step LRA from DP and SDP | Trajectory, BF, and mission time | To maximize accumulated sensing energy and ensure communication QoS. |

This paper [104] explores the joint optimization of resource allocation and trajectory design for a multi-user, multi-target UAV-enabled ISAC system. To align with practical UAV sensing operations, the sensing process occurs while the UAV hovers. The research focuses on optimizing the two-dimensional trajectory, speed, downlink information, and sensing beamformers for a fixed-altitude UAV, with the goal of minimizing average power consumption while meeting the communication quality of service requirements and conducting sensing tasks. To solve the complex non-convex mixed-integer nonlinear programming problem (MINLP), the study applies SDR, the big M method, and SCA to develop an AO algorithm. The simulation results show that this approach delivers substantial power savings compared to two baseline methods that rely on heuristic trajectories.

This paper [105] explores the optimization of the UAV trajectory for simultaneous synthetic aperture radar (SAR) and communication functions, with the objective of minimizing propulsion power while satisfying communication and detection requirements. To tackle the non-convex problem, a trajectory planning algorithm is introduced, which employs SCA and block coordinate descent methods to simplify the process. The simulation results show that the proposed algorithm reduces power consumption by 50% compared to the shortest trajectory method.

The authors in [106] explore ISAC supported by UAVs integrated with mobile edge computing (MEC), where a UAV-mounted ISAC device senses multiple targets and transmits radar data to an edge server for training a machine learning model to recognize these targets. The radar estimation information rate is used to assess sensing performance. The objective is to minimize overall system costs, including UAV energy consumption and data collection time, while satisfying the constraints of model training accuracy and radar sensing performance. A joint optimization problem is formulated that involves the scheduling of signals, the allocation of time slots, the power of the signal and communication, and the trajectory of the UAV. Although the problem is nonconvex, an efficient algorithm is proposed, utilizing vertical decomposition for the layered structure and horizontal decomposition with the block coordinate descent (BCD) method. The numerical results demonstrate the effectiveness of the proposed algorithms and highlight performance improvements.

This paper [107] presents a UAV-enabled ISAC (UAV-ISAC) system, in which the UAV senses ground users and relays the data to a BS. Radar mutual information (MI) is employed to assess the UAV's sensing capabilities. To promote user fairness, the paper addresses a multi-objective resource optimization problem aimed at maximizing both EE and the minimum radar MI. The nonconvex optimization problem is segmented into three subproblems: user scheduling, transmit power, and UAV trajectory optimizations. These are tackled using SCA, fractional programming, and relaxation techniques. By iteratively optimizing each subproblem, a suboptimal solution to the original issue is derived. Simulation results demonstrate that the proposed approach effectively maximizes UAV EE while ensuring fairness among users. In another paper [108], the authors introduce a device-free ISAC system that leverages distributed UAVs to monitor vehicles on urban roads while transmitting data to BSs through wireless communication. To facilitate autonomous UAV decision-making and tackle stochastic optimization challenges with multi-criteria trade-offs in dynamic environments, a MARL algorithm is introduced. Simulation results indicate that this approach significantly improves UAV sensing performance while maintaining stable communication with BSs. Additionally, it promotes fairness in user sensing, enhances UAV flight safety, and reduces energy consumption through hyperparameter tuning.

The authors in [109] introduce an energy-efficient beam alignment and tracking method using ISAC technology to address these issues. Beam alignment relies on predicted angles from the EKF algorithm, adjusting beamwidth by varying the active antenna count based on EKF prediction errors. A block coordinate descent fractional programming algorithm determines the optimal number of active antennas to maximize EE. Simulations demonstrate the effectiveness of this method in improving EE communication.

This paper [110] optimizes energy-sensitive beamforming and trajectory for ISAC using a limited energy UAV. The UAV, with a uniform linear array of half-wavelength dipole antennas, transmits signals to support both downlink communication users and ground target sensing simultaneously. The objective is to maximize sensing energy for ground targets while adhering to energy constraints and ensuring communication quality of service. Achieving this involves joint optimization of the UAV's flight path, ISAC beamforming, and mission duration. The non-convex nature of the problem complicates the pursuit of optimal solutions. In response, a novel, computationally efficient approach is proposed, combining the one-step look-ahead rollout algorithm from approximate dynamic programming (DP) with semidefinite programming from convex optimization. Simulations demonstrate that the proposed method significantly broadens the performance range for both sensing and communication compared to traditional schemes.

**Summary:** This subsection explores EE strategies for ISAC systems enabled by UAVs, focusing on optimization techniques and their limitations. In [103], an ISAC system based on UAVs with WPT improves radar and communication performance using AO, but it assumes ideal energy harvesting, overlooking inefficiencies and variable user demands. Similarly, [108] leverages MARL to enhance sensing and EE in distributed UAV systems, but it faces challenges in dynamic environments with unpredictable patterns. In [104], joint trajectory and beamforming optimization minimizes UAV power consumption, although fixed-altitude operations limit adaptability to 3D scenarios. In terms of addressing beam misalignment, [109] uses EKF-based beam alignment, but performance may degrade with sudden environmental changes. The integration of UAV-MEC in [106] optimizes energy and data collection for machine learning tasks, but it struggles in environments limited by edge servers. Similarly, [107] balances EE and sensing fairness but suffers from computational overhead in large-scale networks.

Energy-aware optimization in [110] maximizes sensing energy but may falter under rapidly changing user demands. The planning of the trajectory of the SAR drone in [105] reduces propulsion power but does not account for real-time environmental factors such as obstacles or weather. In [101], a computational offloading strategy for V2X networks, assisted by UAVs, improves EE but relies on accurate demand prediction, which is challenging under dynamic traffic conditions. Lastly, [102] employs DRL for optimizing the trajectory of the UAV, achieving energy savings but requiring extensive training, which limits real-time scalability. Together, these studies highlight advances in energy-efficient UAV-ISAC systems while underscoring the need for improved adaptability, robustness, and scalability in dynamic environments.

*F. Improving Security*

This subsection explores security enhancements in UAV-enabled ISAC systems, focusing on countering eavesdropping and jamming, as illustrated in Fig. 6. Key strategies include utilizing jammer UAVs to disrupt eavesdroppers and optimizing user scheduling, transmit power, and UAV trajectories to secure data transmission. The integration of RIS can also help to generate artificial noise and improve signal-to-noise ratios, enhancing sensing accuracy and preventing unauthorized access. Additionally, Kalman filtering and trajectory optimization track legitimate users, ensuring secure communication. These approaches, combined with AO and successive convex approximation, significantly improve SRs and overall security in UAV-enabled ISAC systems.

This paper [111] optimizes real-time UAV trajectory design for secure ISAC, with the UAV serving as a downlink transmitter and radar receiver. The legitimate user, Bob, moves through unknown ground locations, whereas the eavesdropper follows a known fixed path. To maximize secrecy rate, an EKF is used to track and predict Bob's position based on delay measurements from echo detection. The trajectory optimization is formulated as a non-convex problem, and an efficient iterative algorithm is proposed to achieve a near-optimal solution. Simulation results indicate that the algorithm

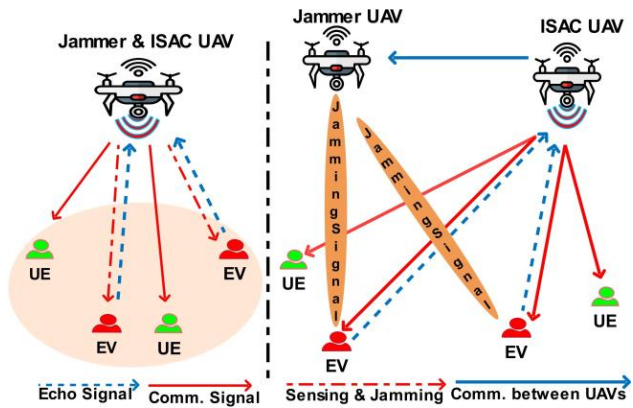

Fig. 6: An illustration of secure UAV-assisted ISAC, where a single ISAC UAV also functions as a jammer (left side), or two separate UAVs coordinate for jamming and ISAC operations (right side).

effectively tracks Bob's movements and balances the trade-off between legitimate and leakage rates.

This paper [112] presents a secure communication system utilizing ISAC-UAVs to counter multiple eavesdroppers. In each time slot, both sensing and communication occur. A jammer UAV disrupts eavesdroppers by broadcasting jamming signals. During the sensing phase, the source UAV detects eavesdroppers and relays their locations to the jammer UAV. While transmitting, eavesdroppers attempt to intercept the source UAV's signals intended for users. The UAV jammer is positioned at the horizontal center of the user group for optimal effectiveness. To enhance secure data transmission, a joint optimization problem is proposed, focusing on user scheduling, transmit power, and the source UAV's trajectory. This non-convex problem is divided into three subproblems, leading to a suboptimal solution via iterative optimization. Simulations demonstrate that the proposed solution significantly increases secure transmission rates despite the presence of multiple eavesdroppers.

This article [113] discusses the benefits of integrating IRS and UAV for ISAC, alongside common security strategies. It proposes two solutions to address jamming and eavesdropping in IRS-UAV-enabled ISAC. For anti-jamming, IRS-UAV creates a line-of-sight link for obstructed targets and employs passive beamforming to mitigate jamming effects. Additionally, the aerial IRS generates artificial noise, improving target sensing accuracy and thwarting eavesdropping. Simulation results validate the effectiveness of these strategies. The article concludes with a discussion on future challenges in this area.

This paper [114] presents a UAV-enabled ISAC system, where a full-duplex UAV with a uniform planar array (UPA) acts as a base station for multiuser downlink communication while simultaneously conducting sensing and jamming against a passive ground eavesdropper. The goal is to maximize the sum secrecy rate for ground users while ensuring sensing accuracy and meeting the UAV's operational constraints. This is achieved through the joint optimization of transceiver beamforming and the UAV's flight trajectory. To solve the problem, a solution using block coordinate descent (BCD) and semidefinite programming (SDP) relaxation is proposed. Simulations verify that the algorithm effectively enhances communication security when sufficient mission time is available.

This paper [115] investigates secure transmission in ISAC systems utilizing IRS and UAV. To counter potential eavesdroppers, artificial noise is introduced to disrupt eavesdropping, boost sensing SNR, and maintain user QoS. A joint optimization approach for UAV deployment, BS transmit beamforming, artificial noise power, and passive beamforming is proposed to maximize the sum secrecy rate. The nonconvex problem is divided into three subproblems and solved using an iterative AO method. Auxiliary variables transform the subproblems into convex forms to address non-convexity. SCA is used for UAV deployment, while semi-definite relaxation determines the optimal BS beamforming, artificial noise power, and passive beamforming. Simulation results demonstrate that the proposed approach significantly enhances ISAC security.

In another paper [116], the authors investigate secure transmission in IRS-assisted UAV-ISAC networks, where the UAV acts as an ISAC BS for K users and J targets via an IRS. A potential Eve attempts to intercept communications without knowledge of the CSI. To counter this, the authors propose a secure transmission strategy that optimizes transmit power allocation, user and target scheduling, IRS phase shifts, and the UAV's trajectory and velocity, thereby enhancing the average achievable rate. An iterative algorithm employing AO, SCA, and manifold optimization (MO) is used for near-optimal solutions, addressing the problem's non-convexity. Additionally, the paper emphasizes energy efficiency maximization with an iterative algorithm that combines AO, SCA, MO, and Dinkelbach's algorithm for non-convex fractional programming. Simulation results validate the effectiveness of these proposed schemes.

**Summary:** The study in [112] enhances secure data transmission by employing a jammer UAV to interfere with eavesdroppers, but its reliance on perfect eavesdropper detection and location accuracy may limit effectiveness in dynamic or complex environments. Similarly, [116] it optimizes UAV trajectory, power allocation, and IRS phase shifts in an IRS-assisted UAV-ISAC network. Yet, the reliance on ideal IRS configurations and complete UAV control makes the approach less robust in scenarios with high interference or resource limitations. In [115], artificial noise and passive beamforming are used to improve secrecy rates in IRS-UAV ISAC systems. Nonetheless, the method's dependence on solving complex iterative optimization problems with auxiliary variables increases computational overhead, which makes it less suitable for real-time or resource-constrained deployments.

The EKF-based trajectory optimization approach in [111] effectively tracks legitimate users to enhance secrecy rates. However, the accuracy of delay measurements used for tracking is highly susceptible to noise and multipath effects. This sensitivity can degrade performance in practical environments characterized by variable signal quality. Meanwhile, the work in [113] integrates IRS and UAVs for ISAC systems to counter jamming and eavesdropping. Finally, the full-duplex UAV-enabled ISAC system in [114] maximizes secrecy rates through joint beamforming and trajectory optimization. Yet, it operates under the assumptions of long mission durations and adequate computational capacity. These requirements may not be feasible given the constraints of battery-powered UAVs or situations that necessitate rapid deployments. Together, these limitations underscore the challenges of implementing secure UAV-ISAC systems in real-world environments that are resource-constrained and subject to dynamic changes.

TABLE VIII: Summary of Secure Communication Schemes for UAV-ISAC Systems

| Ref. | Year | UAV Details | | Communication and Sensing Details | | | | Eve's CSI | Methodology | Opt. Variables | Objective |
|---|---|---|---|---|---|---|---|---|---|---|---|
| | | Role | # (UAVs) | # (UEs) | # (Eves) | Sensing Type | Sensing Metric | | | | |
| [111] | 2023 | T&S | S | S | S | Mobile GUE detection | Delay measurements via sensing echoes | Avail. | SCA, EKF | trajectory | To maximize secrecy rate |
| [112] | 2024 | T&S by jamm. UAV | jamm. & comm. UAV | M | M | Eve detection | Eves' sensing SNR | Avail. | SCA | user scheduling, trajectory, PA | To maximize sum secrecy rate |
| [113] | 2024 | T&S and R&S | S | M | S | Eve detection | - | - | - | - | To maximize sum secrecy rate |
| [114] | 2024 | T&S | S | M | S | Eve sensing | Eve's sensing SNR | Avail. | BCD, SDP | jamming/ sensing BF, information BF, trajectory | To maximize sum secrecy rate |
| [115] | 2024 | R&S | S | M | S | Eve detection | Sensing SNR | Avail. | SCA | UAV position, BF, AN power, RIS PSs | To maximize sum secrecy rate |
| [116] | 2024 | T&S | S | M | S | Target detection | Target's received power | Not avail. | TDMA, SCA, AO, MO | users & targets scheduling, trajectory, PA, RIS PSs | To maximize average secrecy rate |

## IV. LESSONS LEARNED, EMERGING CHALLENGES AND RESEARCH DIRECTIONS

This section outlines key lessons learned from UAV-based ISAC systems and identifies emerging challenges.

### A. Lessons Learned

Integrating UAVs into ISAC provides valuable insights into optimal UAV placements and trajectory design to achieve the desired performance. The following are the key lessons from UAV-integrated ISAC systems:

- **Optimal UAV placement and trajectory:** Optimizing the placement or trajectory of UAV is vital to achieve sum-rate maximization and desired sensing accuracy [76], [79]. By carefully designing the UAV's trajectory or determining its optimal placement, it is possible to enhance coverage, reduce interference, and ensure better alignment with the intended targets.
- **Attention to User Distribution:** Accurately mapping the spatial distribution of users is crucial for UAV trajectory design in UAV-ISAC system [73]. Doing so will help identify the optimal trajectory to maximize the network's overall BS and sensing accuracy.
- **Balancing S&C:** It is crucial to maintain a balance between S&C, as prioritizing one metric can negatively impact the other [88], [93]. Additionally, there is an inherent imbalance in the path loss experienced by S&C signals. This occurs because the sensing signal undergoes greater path loss due to traveling an additional distance after reflecting off the target before reaching the sensing device (another UAV or a BS). This imbalance must be carefully considered in the joint design of S&C.
- **Adaptability to Network Dynamics:** Given the dynamic nature of wireless environments —marked by the movement of both users and sensing targets — UAVs must adapt their trajectory in real-time [76], [81]. The dynamic optimization of UAV-assisted communication and sensing has the potential to significantly enhance overall network functionality by enabling real-time adjustments to varying conditions. This ensures efficient resource allocation, minimizes interference, and improves both sensing accuracy and communication reliability, thereby maximizing the performance of integrated systems.
- **Implementing Strong Security Measures:** A critical takeaway from employing UAVs in ISAC systems is the necessity of robust security measures [111], [115], [116]. It is essential to implement stringent authentication protocols for UAV-based transmission and sensing. Additionally, these systems must be protected from both communication and sensing eavesdroppers. Furthermore, nearby UAVs can also pose a threat by acting as potential eavesdroppers.

### B. Emerging Challenges and Research Directions

- **Joint CE and Optimization:** One of the significant challenges in UAV-ISAC systems is ensuring both accuracy in CE. Inaccuracies in CE, particularly with potential eavesdroppers or jammers, can greatly undermine secrecy rates. Additionally, the continuous motion and hovering-induced jittering of UAVs demand practical system designs [76]. Joint CE and optimization have seen significant advancements. For instance, frameworks such as [117] have enhanced CE accuracy. Additionally, research like [118] has explored optimizing radar probing and UAV trajectories to improve ISAC performance. However, challenges remain, especially in dynamic environments where rapid channel variations and resource

4IEEE INTERNET OF THINGS JOURNAL (FOR REVIEW) 22

- constraints pose difficulties. Future research should focus on developing lightweight, AI-based algorithms that adapt to real-time environmental changes while optimizing CE and resource allocation. Extending optimization frameworks to multi-UAV scenarios by integrating trajectory, beamforming, and resource allocation can further improve system efficiency [119]. Machine learning techniques like reinforcement learning enable UAVs to adapt S&C strategies based on real-time channel conditions, making precise CE and practical system design crucial for successful UAV-ISAC system deployment.
- **Interference Management:** Dynamic spectrum access and interference-resilient waveform designs have made significant strides in interference management. Yet, dense networks and spectrum congestion continue to challenge existing solutions. A good area for research is to look into proactive interference mitigation strategies that use machine learning to guess interference patterns from UAV paths and environmental data [120]. Spectrum sharing in underutilized bands, such as millimeter-wave and terahertz frequencies, could also alleviate congestion. Another area for exploration is swarm-based interference management, where cooperative UAV trajectory and task scheduling can minimize mutual interference while enhancing overall system performance.
- **Mobility Management:** Constantly observing and responding to variations in channel conditions is crucial. This approach enables online trajectory adaptation to achieve better S&C accuracy, rather than relying solely on offline designs [81]. Better trajectory prediction models using machine learning and energy-efficient designs that optimize UAV flight paths have helped with mobility management in UAV-based ISAC systems. Despite these advancements, energy constraints and the complexity of managing large-scale UAV swarms present significant hurdles [56]. Future research can address these challenges by developing energy-efficient mobility strategies that optimize UAV movement while balancing energy consumption with performance. Advanced algorithms for swarm management, focusing on distributed control frameworks, can enhance scalability and operational efficiency. Seamless integration with terrestrial and satellite networks also requires hybrid architectures to ensure consistent connectivity, especially during rapid UAV movements.
- **Security & Privacy:** Quantum cryptography, AI-driven threat detection and RIS-assisted secure system [121], have somehow addressed security and privacy concerns in UAV-based ISAC systems. However, privacy risks associated with inadvertent data collection and cybersecurity threats like hijacking and jamming persist. Future research should explore the implementation of quantum key distribution (QKD) for secure UAV communications, considering their resource constraints. Privacy-preserving algorithms that utilize edge computing frameworks for local data processing can mitigate privacy risks while enhancing operational efficiency. AI-driven systems capable of detecting and countering adversarial attacks also represent a critical area for development [122]. Policy-driven frameworks that ensure compliance with regulations like GDPR while safeguarding data integrity will further support secure and ethical UAV operations.
- **The Need for Standardization:** The lack of standardization in UAV-based ISAC systems limits their interoperability and integration with existing communication infrastructures. Research into globally harmonized protocols and spectrum-sharing standards is necessary to enable seamless interaction between UAVs and terrestrial networks. Developing regulatory policies for dynamic spectrum allocation can optimize resource utilization and reduce interference. Similarly, unified frameworks for UAV operations will facilitate international collaboration and best practices [123].
- **The need for efficient algorithms:** Efficient algorithms are critical for real-time data processing and decision-making in UAV-based ISAC systems [88]. Machine learning has shown the potential to enhance adaptability, but scalability and real-time constraints remain challenging. Research into federated learning can enable distributed optimization, allowing UAVs to collaborate without centralized data aggregation. For UAVs to last longer and make the best use of resources, we need energy-efficient algorithms that keep system performance high while reducing energy consumption [53], [124].
- **Cost and Scalability:** Cost and scalability continue to be significant barriers to the large-scale deployment of UAV-based ISAC systems. Researchers have explored lightweight materials and modular designs, but the high costs of advanced S&C technologies remain prohibitive. Future research should focus on affordable UAV platforms and scalable network architectures. Blockchain-based decentralized resource management systems can also provide innovative solutions for managing large UAV fleets cost-effectively [78].
- **The Impact of Environmental Factors:** Environmental factors such as weather, terrain, and urban interference significantly influence UAV-based ISAC performance [93]. Although researchers have explored adaptive protocols and renewable energy integration, they still face challenges related to signal propagation and sensor accuracy in diverse environments [125]. Research into dynamic environmental adaptation strategies can ensure consistent system performance under varying conditions. Comprehensive impact studies will enhance the advancement of resilient ISAC systems capable at growing under challenging environments.
- **Balancing Optimization and Practicality:** To address UAV limitations, including continuous motion and hovering-induced jittering, a balanced design approach for UAV-based ISAC systems is essential. To address the limitations of UAVs, such as continuous motion and hovering-induced jittering, it is essential to adopt a balanced design approach for UAV-based ISAC systems. It is important to note that a strategy optimized for joint S&C in one time slot may not remain optimal in subsequent slots.
- **Seamless Network Integration** For UAV-based ISAC

systems to function successfully, they must seamlessly integrate with current network protocols and management frameworks. This alignment is essential, as it ensures interoperability and scalability, thereby enhancing the overall functionality within the existing ISAC infrastructure.

- **Communication-Enhanced Sensing:** UAV communication capabilities can enhance sensing efficiency, robustness, and accuracy, complementing the traditional approach of sensing-assisted communication in ISAC systems [125]. UAVs face challenges in computational capacity and meeting low-latency requirements for tasks like target tracking. Offloading intensive sensing tasks to edge servers, such as ground BSs or central UAVs, can address these issues. Efficient resource allocation and selection of nodes with strong LoS links can improve performance, while data compression techniques and multi-hop UAV networks can reduce transmission burdens. Balancing energy consumption with processing delays remains a critical challenge [126].
- **Collaborative Sensing and Fusion:** UAV collaboration enhances sensing accuracy and coverage via shared information, including user positions and environmental changes [92]. This approach reduces resource wastage and facilitates seamless tracking [55]. While data aggregation through a central UAV or BS can deepen target insights, it also introduces latency and resource consumption. Over-the-air computation presents an efficient alternative by aggregating data during transmission, eliminating extra processing steps, and improving fusion efficiency [124].

## V. CONCLUSION

This survey explores UAV-based ISAC systems, emphasizing their pivotal role in 6G and beyond networks. By integrating sensing and communication, UAVs effectively address the challenges posed by traditional systems. They offer critical solutions for various applications, including disaster response, smart cities, and defense operations. The lessons learned underscore the significance of optimizing UAV placement, trajectory design, and balancing SC trade-offs, which are essential for enhancing system performance. However, key challenges remain. In particular, establishing standardized protocols and ensuring seamless integration with terrestrial and satellite networks are crucial for maintaining interoperability and scalability. By addressing these challenges, UAV-based ISAC systems will enable robust, adaptive, and scalable wireless networks, paving the way for transformative applications in the 6G era and beyond. Ultimately, the advancement of these systems will drive further innovation and efficiency across various sectors. By addressing these challenges, UAV-based ISAC systems will enable robust, adaptive, and scalable wireless networks, paving the way for transformative applications in the 6G era and beyond.